\documentclass[12pt]{article}

\usepackage{epsf}
\usepackage{amsmath}
\usepackage{amsfonts}
\usepackage[mathscr]{euscript}
\usepackage{latexsym}

\renewcommand{\theequation}{\thesection.\arabic{equation}} 

\def\scr{\mathscr}
\def\D{{\scr D}}
\def\Db{\bar\D}
\def\Dt{\tilde\D}
\def\Dtb{\bar{\tilde\D}}
\def\L{{\scr L}}
\def\Lb{\bar\L}
\def\C{{\cal C}}
\def\N{{\scr N}}

\def\bR{{\mathbb R}}

\def\e{{\rm e}}
\def\d{{\rm d}}

\newcommand{\sdet}{{\rm sdet}}

\def\pr{\partial}

\newcommand{\gcl}{[\![}
\newcommand{\gcr}{]\!]}

\newcommand{\half}{{\textstyle \frac{1}{2}}}
\def\quar{{\textstyle \frac{1}{4}}}

\newcommand{\Leff}{\L_{\rm eff}}
\newcommand{\Leffb}{\Lb_{\rm eff}}

\newcommand{\Geff}{\Gamma_{\rm eff}}

\newcommand{\dS}{\!\!{\rm d}^6z\,}
\newcommand{\dSb}{\!\!{\rm d}^6\bar z\,}
\newcommand{\dV}{\!\!{\rm d}^8z\,}

\newcommand{\Ikin}{I_{\rm kin}}
\newcommand{\Ig}{I_{g}}
\renewcommand{\Im}{I_{M}}
\newcommand{\Ixi}{I_{\xi}}
\newcommand{\Ii}{I_{1}}
\newcommand{\Iii}{I_{2}}

\newcommand{\Lkin}{\L_{\rm kin}}
\newcommand{\Lg}{\L_{g}}
\newcommand{\Lm}{\L_{M}}
\newcommand{\Lxi}{\L_{\xi}}
\newcommand{\Lxib}{\L_{\xi}'}
\newcommand{\Li}{\L_{1}}
\newcommand{\Lib}{\L_{1}'}
\newcommand{\Lii}{\L_{2}}
\newcommand{\Liib}{\L_{2}'}
\newcommand{\Lbox}{\L_{\Box}}

\newcommand{\ukin}{u_{\rm kin}}
\newcommand{\ug}{u_{g}}

\newcommand{\uxi}{u_{\xi}}
\newcommand{\uxib}{u_{\xi}'}
\newcommand{\ui}{u_{1}}
\newcommand{\uib}{u_{1}'}
\newcommand{\uii}{u_{2}}
\newcommand{\uiib}{u_{2}'}
\newcommand{\ubox}{u_{\Box}}

\newcommand{\al}{\alpha}
\def\da{{\dot\alpha}}
\def\be{\beta}
\def\db{{\dot\beta}}
\def\dg{{\dot\gamma}}
\def\bt{{\bar \theta}}
\def\is{{^{\!(\sigma)}}}
\def\il{{^{\!(\Lambda)}}}
\def\de{\delta}
\def\si{\sigma}

\newcommand{\eps}{\varepsilon}

\newcommand{\tfr}[2]{{\textstyle \frac{#1}{#2}}}

\newcommand{\fdq}[2]{\frac{\delta #1}{\delta #2}}

\newcommand{\tf}[1]{$\displaystyle #1 $}

\def\ts{\textstyle}

\begin{document}
\thispagestyle{empty}
\begin{flushright}
hep-th/9804053\\ NTZ 7/98\\ April 1998
\end{flushright}
\begin{center}
{\Large 
 \bf Conformal Transformation Properties of the }

{\Large \bf Supercurrent in Four  Dimensional}  

{\Large \bf Supersymmetric Theories }

\vspace{0.7cm}

{\parindent0cm
Johanna Erdmenger\footnote{Supported by Deutsche Forschungsgemeinschaft,
e-mail: Johanna.Erdmenger@itp.uni-leipzig.de}\\
Christian Rupp\footnote{Supported by Graduiertenkolleg
  "Quantenfeldtheorie: Mathematische Struktur und physikalische
Anwendungen",
e-mail: Christian.Rupp@itp.uni-leipzig.de}\\
Klaus Sibold\footnote{e-mail: sibold@physik.uni-leipzig.de}
}

\vspace{1cm}

Institut f{\"u}r Theoretische Physik\\
Universit{\"a}t Leipzig\\
Augustusplatz 10/11\\
D - 04109 Leipzig\\
Germany
\end{center}

\vspace{1cm}

\centerline{\small \bf Abstract}\vspace*{-2mm} { \small We investigate
the superconformal transformation properties of Green functions with
one or more insertions of the supercurrent in $N=1$ supersymmetric
quantum field theories. These Green functions are conveniently
obtained by coupling the supercurrent and its trace to a classical
supergravity background. We derive flat space superconformal Ward
identities from diffeomorphisms and Weyl transformations on curved
superspace. For the classification of potential quantum superconformal
anomalies in the massless Wess-Zumino model on curved superspace a
perturbative approach is pursued, using the BPHZ scheme for
renormalisation.  By deriving a local Callan-Symanzik equation the
usual dilatational anomalies are identified and it is shown that no
further superconformal anomalies involving the dynamical fields are
present.}

\vspace*{5mm}
\begin{tabbing}
\> Curved Superspace
Background, Supercurrent, Anomalies.
\end{tabbing}
\newpage

\section{Introduction}

Conformal symmetry is a topic which is of interest from a multitude of
points of view and which therefore has received attention repeatedly
throughout the years. Of parti\-cular interest is superconformal
symmetry which links conformal space-time symmetry to an internal
symmetry.  Supersymmetric quantum field theories provide many
non-trivial examples for four dimensional theories with vanishing beta
functions. Such theories are expected to be conformally invariant on
flat space. Within perturbation theory, necessary and sufficient
conditions for the beta functions vanishing to all orders in
supersymmetric gauge theories have been given and used for the
construction of finite theories in which the beta functions vanish
identically \cite{LPS}.

Within the context of electric-magnetic duality, there are examples
for $N=1$ supersymmetric gauge theories with non-trivial
renormalisation group fixed points \cite{Seiberg}.  These theories may
be used for testing conjectured extensions of the Zamolodchikov C
theorem \cite{Za} to four dimensions.  Recently several examples of
$N=1$ supersymmetric Yang-Mills theories have been explored
\cite{Anselmi} in which the coefficient of the topological Euler
density is larger in the UV than in the IR limit, in agreement with a
possible four dimensional analogue of the C theorem.

In this paper we consider supersymmetric theories away from the fixed
points, where anomalies involving the dynamical fields may be
present. It is appropriate to use the superfield formalism in which
supersymmetry is manifest. We study the transformation properties of
the supercurrent under superconformal transformations.  The
supercurrent is an axial vector superfield which has the $R$ and
supersymmetry currents among its components, as well as the energy
momentum tensor.  Its space-time moments and derivatives yield all the
currents of the superconformal group.  The superconformal
transformation properties of Green functions for elementary fields are
well-known \cite{PS78} and require considering Green functions with
one insertion of the supercurrent only. Here however, we derive the
transformation properties of Green functions with multiple insertions
of the supercurrent. In analogy to the original proof of the C theorem
in two dimensions, these are expected to be relevant for a possible
proof of a C theorem for four dimensional supersymmetric theories.

Multiple insertions of the supercurrent are conveniently generated by
coupling the quantum field theory considered to a classical curved
superspace background.  The constraints on torsion and curvature are
chosen such as to obtain minimal supergravity with a real vector
superfield prepotential, to which the supercurrent couples, and with a
chiral compensator. Superconformal Ward identities are obtained by
combining diffeomorphisms and Weyl transformations on curved
superspace. For definiteness we consider the massless Wess-Zumino
model, which is the simplest example for a classically superconformal
theory. We expect to extend our results to gauge theories in the
future.

For the quantisation of the dynamical fields we adopt a perturbative
approach analogous to the method employed in \cite{KS3,KS1,KS2} in a
related study of Green functions with double insertions of the energy
momentum tensor in scalar $\phi^4$ theory.  This approach ensures an
off-shell formulation at every stage of the analysis and yields
scheme-independent results which hold to all orders in perturbation
theory. Thus all potential anomalies are systematically investigated.
At the operator level, the Ward identities for Green functions with
multiple insertions of the supercurrent constitute the current algebra
for all currents of the superconformal group including its anomalies.

In section 2 of this paper we review the elements of classical
supergravity which are necessary for our analysis \cite{bk,ggrs}.
Furthermore we discuss the superconformal transformation properties of
the supergravity fields. In section 3 we discuss superconformal Ward
identities for classical $N=1$ supersymmetric field theories on curved
superspace and apply them in particular to the Wess-Zumino model. We
show how the Ward identities derived on flat space in \cite{PS78} (for
a review see \cite{PS}) follow naturally from the superconformal
transformations on curved superspace discussed in section 2. These
Ward identities lead to the definition of the supercurrent and of its
trace.  In section 4 we quantise the Wess-Zumino model on curved
superspace using the BPHZ scheme for renormalisation. Green functions
with insertions of the supercurrent are defined.  The conditions for
the validity of a Callan-Symanzik equation are given.  With the help
of symmetry consistency conditions \cite{O} we are able to show that
symmetry breaking terms of matter-background coupling type may be
removed in agreement with renormalisability. Furthermore we derive a
local Callan-Symanzik equation which determines the anomaly structure
of the full superconformal group. In section \ref{sectransprop} we discuss the
transformation properties of Green functions with supercurrent
insertions which follow from the the local Callan-Symanzik equation.
Section \ref{secdisc} contains a discussion as well as some concluding remarks.
Some longer formulae are relegated to the appendix.

\section{Superconformal Transformations in Curved Superspace}

\setcounter{equation}{0}

We begin this section by presenting a few well-known results of 
classical supergravity and by introducing our conventions. 
Subsequently these facts enable us
to discuss diffeomorphisms and Weyl transformations in curved superspace.
The combination of these two transformations is shown to reduce to
the superconformal transformations when re\-stricting to flat superspace.

We consider the superspace manifold $\bR^{4|4}$. At each point $z =
(z^M) = (x^m, \theta^\mu,\bar\theta_{\dot\mu}) \in \bR^{4|4}$, the
diffe\-rential operators
$(\partial_M)=(\partial_m,\partial_\mu,\bar\partial^{\dot\mu})$ span
the tangent space. A supervierbein is a set of eight vector fields
\begin{equation}
(\tilde E_A) = (\tilde E_a, \, \tilde E_\alpha, \, \bar {\!\tilde
  E}{}^{\dot\alpha}),
\end{equation}
such that the $\tilde E_A(z)$ constitute a basis of the tangent space
at each point $z$.  The supervierbein can be regarded as the gauge
covariant derivative for the gauge group of diffeomorphisms in
superspace.  We require supervierbeins related to each other by a
superlocal Lorentz transformation to be physically equivalent, such
that the gauge group is extended to the direct product of the group of
diffeomorphisms and the superlocal Lorentz group with generators
$M_{\alpha\beta}$, $\bar M_{\dot\alpha\dot\beta}$ acting in tangent
space.  The extended gauge group has an additional gauge field $\tilde
\omega_A^{\alpha\beta}$ , such that the gauge covariant derivative is
given by
\begin{equation}
\Dt_A = \tilde E_A + \tilde \omega_A.
\end{equation}
Under diffeomorphisms $K$ and Lorentz transformations $M=
K^{\alpha\beta} M_{\alpha\beta}+\bar K^{\dot\alpha\dot\beta} \bar
M_{\dot\alpha\dot\beta}$ these derivatives change in the following
way:
\begin{equation}
 \Dt_A \longrightarrow \Dt_A' = \e^K \, \Dt_A \,\e^{-K} \, ,
 \hspace{0.7cm}
 \label{Ktrans}
 \Dt_A \longrightarrow \Dt_A' = \e^M \, \Dt_A \,\e^{-M}.
\end{equation}
Torsion and curvature of superspace are obtained from the graded
commutator of covariant derivatives,
\begin{equation}
 \gcl \Dt_A, \Dt_B \gcr = {T_{AB}}^C \Dt_C + {R_{AB}}^{\alpha\beta}
 M_{\alpha\beta} + {R_{AB}}^{\dot\alpha\dot\beta} \bar
 M_{\dot\alpha\dot\beta}.
\label{gradcomm}
\end{equation}
So far we have used the operators $\partial_A$ as a basis in tangent
space.  An alternative basis is given by the flat covariant
derivatives
\begin{eqnarray}
(D_A) = ( D_a, D_\alpha, \bar D^{\dot\alpha} ), \hspace{4cm} \\ D_a =
\partial_a, \;\;\;\; D_\alpha = \partial_\alpha -i
\sigma^a_{\alpha\dot\alpha} \bar \theta^{\dot \alpha} \partial_a,
\;\;\;\; \bar D_{\dot\alpha} = -\partial_{\dot\alpha} +i \theta^\alpha
\sigma^a_{\alpha\dot\alpha} \partial_a. \nonumber
\end{eqnarray}
The spinor covariant derivatives satisfy the anticommutation relation
\begin{align}
\Big\{ D_\al, \bar D_\da \Big\} &= 2 i \sigma^b_{\al \da} \pr_b \, .
\end{align}
We may also define $D_{\alpha\dot\alpha} = \frac{1}{2i}
\{D_\alpha,\bar D_{\dot\alpha}\}$ which replaces $D_a$.  Using this
basis, the vierbein is given by
\[\tilde E_A = \tilde E_A{}^M D_M = \tilde E_A{}^{\mu\dot\mu} D_{\mu\dot\mu} 
+ \tilde E_A{}^\mu D_\mu + \bar {\!\tilde E}_{A\dot\mu} \bar
D^{\dot\mu}.\]

In order to be able to introduce covariantly chiral scalar
superfields, it is necessary to impose constraints on the torsion and
curvature components. In this paper we impose the torsion constraints
of conformal supergravity and the additional condition $T_\al=0$,
$T_\al = T_{\al b}{}^b - T_{\al \be}{}^\be- T_{\al \db}{}^\db$ with
$T_{AB}{}^C$ defined in (\ref{gradcomm}), such that the superconformal
group is the maximal symmetry group.  Some of these constraints
determine $\omega$ in terms of the vierbein, while the remaining can
be solved by introducing prepotentials $W$ and $F$ such that the
covariant derivatives may be expressed by
\begin{eqnarray}
\Dt_\alpha = \e^W (F\, D_\alpha) \e^{-W} \, +\, \tilde
\omega_\alpha,\;\;\;\; \Dtb_{\dot\alpha} = \e^{\bar W} (\bar F\, \bar
D_{\dot \alpha}) \e^{-\bar W} \, +\, \bar {\tilde \omega}_{\dot
\alpha}, \nonumber\\ \Dt_a = -\frac{i}{4}
{\sigma_a}^{\alpha\dot\alpha} \{ \Dt_\alpha, \Dtb_{\dot\alpha} \},
\;\;\;\; W = W^A D_A + W^{\alpha\beta} M_{\alpha\beta} +
W^{\dot\alpha\dot\beta} \bar M_{\dot\alpha\dot\beta},
\label{WFdef}
\end{eqnarray}
where $W^{\alpha\beta}$ and $W^{\dot\alpha\dot\beta}$ are symmetric,
and $W^A$, $W^{\alpha\beta}$, $W^{\dot\alpha\dot\beta}$ and $F$ are
complex superfields.  $F$ is constrained by
\begin{equation}
 \bar {\!\tilde E}_{\dot\alpha} \, \left( \sdet ({\tilde E_A}^M) \,
 \bar F^{2} \, (1\cdot \e^{\overleftarrow{W}}) \right)
 =0. \label{F_constraint}
\end{equation}
\hspace{\parindent}The prepotentials determining the covariant
derivatives in \eqref{WFdef} are not unique.  There is some
arbitrariness in $W$ and $F$ which manifests itself in the so-called
$\Lambda$ group of transformations.  This group is the group of
transformations $\Lambda$ which leave the covariant
derivatives invariant,
\begin{equation}
\e^\Lambda \, \Dt_A \, \e^{-\Lambda} = \Dt_A,
\label{lambdatrans}\;\;\;\; \Lambda = \Lambda^A D_A + \Lambda
^{\alpha\beta} M_{\alpha\beta} + \Lambda^{\dot\alpha\dot\beta}
M_{\dot\alpha\dot\beta}.
\end{equation}
Thus the gauge group is enlarged further and consists of
diffeomorphisms, superlocal Lorentz transformations and $\Lambda$
transformations.  This situation can be simplified by using the
chiral representation
\begin{equation}
\D_a = \e^{-\bar W} \Dt_a \e^{\bar W}, \qquad \Db_\da = \e^{-\bar W}
\Dtb_\da \e^{\bar W}, \qquad A = \e^{-\bar W} \tilde A\, .
\end{equation}
The chiral scalar field $A$ plays the role of the matter field in the
following sections.  This representation is analogous to the flat
superspace chiral representation in which all fields $\tilde \Phi$ are
replaced by $\Phi = \e^{i\theta\sigma^a\bar\theta\partial_a}
\tilde\Phi$.  Furthermore we define the operator $H$ by
\begin{equation}
\e^{2i H } = \e^{-\bar W} \e^{W},
\end{equation}
where $H = H^A D_A + H^{\alpha\beta} M_{\alpha\beta} + \bar
H^{\dot\alpha\dot\beta} M_{\dot\alpha\dot\beta}$ is a real superfield.
As in the flat space chiral representation, the complex conjugate of
an expression in the chiral representation belongs to the antichiral
representation. Whenever quantities in the chiral and antichiral
representation are combined into the same expression, a representation
changing factor $\e^{2iH}$ has to be included. For example, $\tilde A
\bar{\!{\tilde A}\,}$ becomes $A \e^{2iH}\bar A$ in the chiral
representation.  It should also be noted that in our conventions
$\Db_\da$ is in the chiral and $\D_\al$ is in the antichiral
representation.

In the chiral representation the $K$ and $M$ transformations are
eliminated and the equations (\ref{Ktrans}) are replaced by
\begin{equation}
\D_A \longrightarrow \e^\Lambda \D_A \e^{-\Lambda} \, , \;\;\;\;
\e^{2iH} \longrightarrow \e^\Lambda \e^{2iH} \e^{-\bar \Lambda},
\end{equation}
where
\begin{eqnarray}
\Lambda = \Lambda^A D_A + \Lambda ^{\alpha\beta} M_{\alpha\beta} +
\Lambda^{\dot\alpha\dot\beta} M_{\dot\alpha\dot\beta}, \nonumber\\
\bar D_{\dot\beta}\Lambda^{\alpha\dot\alpha} = 4i \Lambda^\alpha
\delta_{\dot\beta}^{\dot\alpha}, \label{lambdarestrict_a}\;\;\;\; \bar
D_{\db} \Lambda^\alpha =0, \;\;\;\; \Lambda_{\dot\alpha\dot\beta} =
-\bar D_{(\dot\alpha} \Lambda_{\dot\beta)}.
\end{eqnarray}
In the chiral representation the $\Lambda $ transformations play thus
the role of (restricted) diffeomorphisms and Lorentz transformations.
Moreover it should be noted that $\Lambda$ is complex. 
Its conjugate $\bar \Lambda$ is constrained by relations which are obtained
from (\ref{lambdarestrict_a}) by complex conjugation.

Some of the $\Lambda$ transformations may be used to restrict $H$ to
the simpler form
\begin{equation}
H = H^{\alpha\dot\alpha} D_{\alpha\dot\alpha} ,
\end{equation}
which imposes the conditions
\begin{equation}
\Lambda_{\dot\alpha}= \e^{2iH} \bar \Lambda_{\dot\alpha} \, , \;\;\;\;
\Lambda_{\alpha\beta} = \e^{2iH} \bar \Lambda_{\alpha\beta}\, .
\label{lambdarestrict_e}
\end{equation}
The Lorentz generators $M_{\alpha\beta}$ and $\bar
M_{\dot\alpha\dot\beta}$ act only on fields which carry spinor
indices. Since we are not going to consider such fields, we set them
to zero from now on.  The $\Lambda$ transformations are then simply
\begin{equation} 
\Lambda = \Lambda^{\alpha\dot\alpha} D_{\alpha\dot\alpha} +
\Lambda^{\alpha}D_\alpha + \Lambda_{\dot\alpha} \bar D^{\dot\alpha} .
\end{equation}

The restrictions (\ref{lambdarestrict_a}) may be solved in terms of a
complex superfield $\Omega^\alpha$,
\begin{equation}
\Lambda^{\alpha\dot\alpha} = i \bar D^{\dot\alpha} \Omega^\alpha\,
,\;\;\;\; \Lambda^\alpha = \frac{1}{4} \bar D^2 \Omega^\alpha \, .
\label{omegadef}
\end{equation}
Equation \eqref{omegadef} defines $\Omega^\alpha$ uniquely up to a
chiral field. Similarly $\bar \Lambda^{\al \da}, \bar \Lambda_\da$ are
determined in terms of $\bar \Omega^\da$. (\ref{lambdarestrict_e}) and
its conjugate ensure that $\Lambda_\da$ is a function of $\bar \Omega_\da$
and $\bar \Lambda^\al$ is a function of $\Omega^\al$.

An important quantity for the construction of invariant actions is the
determinant of the supervierbein, $E\equiv \sdet({E_A}^M)$.
Constraint \eqref{F_constraint} now simply reads
\[\bar D_{\dot\alpha} \, (E \bar F^2)  =0.\]
This can easily be solved by setting
\begin{equation}
E \bar F^2 = \phi^{-3},
\label{phidef}
\end{equation}
where the chiral compensator $\phi$ satisfies $\bar{D}_{\da} \phi =0$.

Furthermore there are super Weyl transformations $\sigma,
\bar{\sigma}$.  If $\bar{D}_{\da} \sigma =0$ and ${D}_{\alpha} \bar
\sigma =0$, the Weyl transformations leave the (anti-) commutation
relations of the covariant derivatives unchanged and respect the
constraints on torsion and curvature.  For the Weyl transformation
properties of the prepotentials we have
\begin{equation}
H \longrightarrow H,\;\;\;\; \phi \longrightarrow \e^\sigma \phi.
\end{equation} 
For our subsequent investigations it is crucial to note that $H$ is a
Weyl invariant.  The transformation laws of the prepotentials and of a
chiral matter field $A$ under $\Lambda$ and super Weyl transformations
are summarised in appendix \ref{transformationlaws}.

With the help of the prepotential $\bar{F}$ we may define the
supersymmetric curvature scalar $R$ which in the chiral representation
is given by
\begin{gather}
R = \bar{D}^2 \, \bar{F}^2 \, .
\end{gather}
The transformation properties of $\bar{F}$ and $F$ imply that $R$ and
$\bar R$ transform homogeneously under $\Lambda$ transformations. The
Weyl transformation properties of $R$, $\bar R$ are given by
\begin{gather}
\delta_\sigma R \, = \, - 2 \sigma R, \;\;\;\; \delta_{\bar{\sigma}} R
\, = \, (\Db^2+R) \bar\sigma \, , \qquad \delta_{\bar \sigma} {\bar R}
\, = \, - 2 {\bar \sigma} \bar{R} , \;\;\;\; \delta_{{\sigma}} {\bar
R} \, = \, (\D^2+\bar R) \sigma \, .
\end{gather}

Moreover, $R$ and $\bar R$ are essential for defining a chiral
(antichiral) integration measure by
\begin{gather} \label{chiralint}
\int \! d^{\, 8} z \, E^{-1} \;\;\; = \;\;\; \int \! d^{\, 6} z \,
\phi^3 \left( \Db^2 + R \right) \;\;\; = \;\;\; \int \! d^{\, 6} {\bar
z} \, \bar \phi^3 \left( {\D}^2 + \bar R \right) \e^{-2iH} \, .
\end{gather}
The factor $\e^{-2iH}$ ensures the change from the chiral to the
antichiral representation.

We proceed by defining infinitesimal transformations which combine
diffeomorphism ($\Lambda$) and Weyl transformations.  The
infinitesimal form of diffeomorphisms and Weyl transformations is
obtained by expanding the expressions of appendix
\ref{transformationlaws}.  For the fields $H_{\alpha \da}$, $\phi$ and
$A$, the combined infinitesimal transformations are given by
\begin{gather} \label{conftrans}
\delta H_{\alpha \da} = \delta_\Lambda H_{\alpha \da} \, , \qquad
\delta \phi = \delta_\Lambda \phi + \delta_\sigma \phi \, , \qquad
\delta A = \delta_\Lambda A + \delta_\sigma A \, ,
\end{gather}
where
\begin{gather}
2i \delta_\Lambda H = \Lambda-\bar\Lambda - {\ts\half} [2iH,
\Lambda+\bar \Lambda] +{\ts\frac{1}{12}} [2iH, \left[2iH, \Lambda-\bar
\Lambda] \right] + O(H^3) \, , \nonumber\\
\label{delta_Lambda_H}
\delta_\Lambda \phi = \Lambda \phi + {\textstyle \frac{1}{3}} \left(
D^{\alpha \da} \Lambda_{\alpha \da} + D^\alpha \Lambda_\alpha \right) \phi
\, , \qquad \delta_\Lambda A = \Lambda A \, , \qquad \\ \delta_\sigma
\phi = \sigma \phi \, , \qquad \delta_\sigma A = - \sigma A
 \label{delta_sigma_AJH} \, .
\end{gather}
We note that the $\Lambda$ transformation properties of $H$ are
analogous to the gauge transformations of a non-abelian gauge field.
For later convenience it is useful to define a chiral dimensionless
field $J$ by
\begin{align}
\phi &\equiv \e^J = 1 + J + \half J^2 + \dots \, ,
\end{align}
such that flat superspace is characterised by vanishing external
fields $H=0, J=0$ rather than by $H=0, \phi=1$. From its definition
and the transformation properties of $\phi$ the infinitesimal
transformations of $J$ are given by
\begin{gather}
\delta J = \Lambda J + {\ts \frac{1}{3}} (D^{\alpha \da} \Lambda_{\al
\da} + D^\al \Lambda_\al ) + \sigma \, .
\end{gather}

To lowest order in an expansion around flat space, $ H $ and $ J $ remain
invariant if
\begin{subequations}
\begin{gather} 
\Lambda =\bar\Lambda \, , \label{Lflach} \\
\label{sigmaflach}
\sigma = -{\ts\frac{1}{3}} \left(D^{\al\da} \Lambda_{\al\da}+ D^\al
\Lambda_\al \right) = - {\ts \frac{1}{12}} \bar{D}^2 D^\alpha
\Omega_\alpha \, .
\end{gather}
\end{subequations}
Together with the constraints (\ref{lambdarestrict_a}, \ref{lambdarestrict_e}),
the condition (\ref{Lflach}) is a necessary and sufficient condition
for $\Omega^\al$ to be restricted to the superconformal coordinate
transformations.
The solutions to (\ref{Lflach}) are listed in appendix 
\ref{superconformalparameters}. To obtain the superconformal
transformation properties of fields on flat space, it is necessary
to combine their infinitesimal diffeomorphism and Weyl transformations
and to impose both (\ref{Lflach}) and (\ref{sigmaflach}).

It should be noted
that on curved space the use of (\ref{Lflach}) is not mandatory even if 
the Weyl transformations are restricted by (\ref{sigmaflach}). Throughout
this paper (\ref{Lflach}) is imposed only when the flat space limit
is taken. 
To obtain the combined  diffeomorphism and Weyl transformations
of the fields $H,J,A$ in an expansion around flat
 space, it is convenient to impose
(\ref{sigmaflach}) even on curved space
in view of expressing the combined transformations
with  the superfields $\Omega, \bar{\Omega}$ as their only parameters.
In this form the infinitesimal combined  transformations are listed in
appendix \ref{combinedtransformations}.

\section{Classical Theory}

\setcounter{equation}{0}

\subsection{Ward Identities and Supercurrent}

The superconformal group is generated by translations, supersymmetry,
$R$ and Lorentz transformations, dilatations, special conformal and
special supersymmetry transformations. The action of the generators
$G$ belonging to these transformations on a chiral field $A$ is given
by
\begin{equation} \label{algebra}
\left[ i G \, , \, A(z) \right] = \delta^G A(z) \, .
\end{equation}
The explicit form of the generators and the corresponding
infinitesimal transformations belong to the standard results of
supersymmetry theory and may be found in textbooks \cite{PS}. The
generators $R, Q_\al, \bar{Q}^\da$ and $P^a$ of $R$ and supersymmetry
transformations and of translations form a subalgebra of the full
superconformal algebra. The currents associated with these four
generators form a supermultiplet, the supercurrent.

The flat space superconformal transformation properties of a flat
space field theory with chiral and antichiral superfields $A, \bar{A}$
are characterised by the local Ward identity
\begin{gather}
W_\Omega \hat \Gamma = - i \int \! \d^6 z \; \delta_\Omega A
\frac{\delta \hat \Gamma}{ \delta A} \, - \, i \, \int \! \d^6 \bar{z}
\; \delta_{\bar{\Omega}} \bar{A} \frac{ \delta \hat \Gamma}{ \delta
\bar{A}} \; ,
\end{gather}
where $\Omega$ is defined in (\ref{omegadef}), $\bar{\Omega}$ is its
conjugate field and $\hat\Gamma$ the classical action.  From the
infinitesimal conformal transformation $\delta A$ in (\ref{conftrans})
we have in the flat space case using (\ref{sigmaflach})
\begin{gather} \label{delta_Omega_A} 
\delta_\Omega A = \bar D^2 \left( {\ts \frac{1}{4}} \Omega^\alpha
D_\alpha A \, +{\ts \frac{1}{12}} D^\alpha \Omega_\alpha A \right) \,
, \qquad \delta_{\bar{\Omega}} \bar{A} = D^2 \left( {\ts \frac{1}{4}}
\bar{\Omega}_\da \bar{D}^\da \bar{A} \, + {\ts \frac{1}{12}}
\bar{D}_\da \bar{\Omega}^\da \bar{A} \right) \, ,
\end{gather}
and thus
\begin{gather} \label{AWI}
W_\Omega \hat \Gamma \, = \, - \, i \, \int \! \d^8 z \, \left(
\Omega^\alpha w_\alpha + \bar{\Omega}_{\da} \bar{w}^{\da} \right) \,
\hat \Gamma
\end{gather}
with
\begin{gather}
w_\al = \quar D_\alpha A \fdq{}{A} -{\ts \frac{1}{12}} D_\alpha \left(
A \fdq{}{A} \right) \, , \qquad \bar{w}_\da = \quar \bar{D}_{\da}
\bar{A} \fdq{}{\bar{A}} -{\ts \frac{1}{12}} \bar{D}_\da \left( \bar{A}
\fdq{}{\bar{A}} \right) \, .
\end{gather}

By comparing with the superconformal transformation law
(\ref{algebra}) we may identify the components of $\Omega$ which lead
to the different superconformal transformations. These are listed in
appendix \ref{superconformalparameters}.

From appendix \ref{superconformalparameters} we read off that for the
$R$, $Q$, $\bar{Q}$ and $P$ transformations we may write
\begin{gather}
\Omega_\al = D_\al \omega \, , \qquad \bar{\Omega}_\da = - \bar{D}_\da
\omega
\end{gather}
with $\omega$ an imaginary superfunction given by
\begin{align}
\omega & = \half i \theta \sigma \bt t - \bt^2 \theta \cdot q +
\theta^2 \bt \cdot \bar{q} - \half i \theta^2 \bt^2 r \\ &= -8 (ir +
q^\al \pr_\al + \bar{q}_\da \bar{\pr}^\da + \quar i \sigma^a_{\al \da}
t_a \pr^\al \bar{\pr}^\da ) \, \delta^{(2)}(\theta) \delta^{(2)}(\bt)
\, , \label{omegadelta}
\end{align}
with $\delta^{(2)}$ the delta function.  Thus for the $R,Q,\bar{Q},P$
transformations we have
\begin{align}
-i \int \dV \left( \Omega^{X \alpha} w_\alpha +
\bar{\Omega}^{X}{}_{\da} \bar{w}^{\da} \right) \, \hat \Gamma & = \, i
\, \int \dV \omega \, \left( D^{ \alpha} w_\alpha - \bar{D}_{\da}
\bar{w}^{\da} \right) \, \hat \Gamma \, \nonumber\\ & \equiv \, i \,
\int \dV \omega \, w \, \hat \Gamma \, , \label{wdef2}
\end{align}
for which we find, using the expression (\ref{omegadelta}) for
$\omega$, the component decomposition
\begin{align}
i \, \int \! \d^8 z \, \omega \, w \, &= r W^R + q^\al W^Q_\al +
\bar{q}_\da W^{\bar Q \da} + t^a W^P_\al
\end{align}
with
\begin{align}
W^R \hat \Gamma &= \, 8 \, \int \!\! \d^{\,4} x \,w \, \hat\Gamma
\Big|_{\theta=\bt=0} \, \qquad W^Q_\al \hat \Gamma = \, 8 i \pr_\al \,
\int \!\! \d^{\,4}x \, w \, \hat\Gamma \Big|_{\theta=\bt=0} \, ,
\nonumber \\ W^{\bar Q\da} \hat\Gamma &= \, 8 i \bar{\pr}^\da \, \int
\!\! \d^{\,4}x \, w \, \hat\Gamma \Big|_{\theta=\bt=0} \, \qquad
W^{Pa} \hat\Gamma = \, 2 \, \sigma^a_{\al \da} \pr^\al \pr^\da \, \int
\!\! \d^{\,4}x \, w \, \hat\Gamma \Big|_{\theta=\bt=0} \, .
\end{align}
Then the operator
\begin{align} \label{wdef}
{\hat W}^R &= \, 8 \, \int \!\! \d^{\,4}x \, w
\end{align}
is given by
\begin{eqnarray}
{\hat W}^R = W^R - i\theta W^Q + i \bt W^{\bar Q} - 2 \theta \si^\mu
\bt \, W_\mu{}^P \, .
\end{eqnarray}

For a superconformal theory for which
\begin{equation}
\hat{W}^R \hat\Gamma = 0 \, ,
\end{equation}
the integrand is a total divergence
\begin{equation} \label{w}
w \, \hat\Gamma = -i {\ts \frac{1}{8}} \, \pr^a V_a \, ,
\end{equation}
with $V_a$ a (axial-)vector superfield.  Allowing also for symmetry
breaking terms, (\ref{w}) may be decomposed
into a chiral and an antichiral equation using (\ref{wdef2}),
\begin{gather} \label{traceequation}
 - 16 w_\alpha \hat \Gamma = \bar D^\da V_{\alpha \da} \, + \, 2 \,
D_\al S - B_\alpha \, , \qquad - 16 \bar w_\da \hat \Gamma = D^\al
V_{\alpha \da} \,+ \, 2\, \bar D_\da \bar{S}\, - \, \bar B_\da \, , \\
V_{\alpha \da} = \half \si^\mu{}_{\alpha \da} V_{\mu} \, \qquad V_a =
\si_a{}^{\alpha \da} V_{\alpha \da} \, , \nonumber
\end{gather}
with $B_\alpha$ satisfying the constraint
\begin{equation}
D^\alpha B_\alpha - \bar D_\da \bar B^\da = 0 \, .
\end{equation}
(\ref{traceequation}) is referred to as the trace identity.  It leads
to the Ward identity
\begin{align} \label{sc}
\pr^a V_a &= 8 i w \hat\Gamma + i ( D^2 S - \bar D^2 \bar S) \, .
\end{align}
Among the components of the supercurrent there are the $R$ and
supersymmetry currents as well as the energy momentum tensor.
However, as is well known \cite{PS78},\cite{PS}, these currents are
not conserved if symmetry breaking terms of both S and B type are
present simultaneously.  The construction of conserved currents
differs according to which type of breaking is present.

In massless theories, the S and B formulations are both possible.  In
massive theories, only the S formulation is consistent since mass
terms are of this type.  Since we would like to understand massless
theories as the well-defined limit of massive theories, we consider
only S type breaking terms in the following.  The trace equations then
read
\begin{gather} \label{traceidentity}
 - 16 w_\alpha \hat\Gamma = \bar D^\da V_{\alpha \da} \, + \, 2 \,
D_\al S \, , \qquad - 16 \bar w_\da \hat\Gamma = D^\al V_{\alpha \da}
\,+ \, 2\, \bar D_\da \bar{S}\, ,
\end{gather}
and the identification of $\theta$-components of $V_a$ with component
currents is such that the superfields
\begin{align}
\hat{R}_a &= V_a \, ,\nonumber\\ \hat{Q}_{a \al} &= i \left( D_\al V_a
- (\sigma_a \bar{\sigma}^b)_\al{}^\beta D_\beta V_b \right)
\nonumber\\ \hat{T}_{a b} &= - {\ts \frac{1}{4}} \left( V_{ab} +
V_{ba} \right) + \half g_{ab} V_\lambda{}^\lambda \, \nonumber\\ &
\hspace{2cm} V_{ab} = {\ts \frac{1}{4}} \left( D \sigma_a \bar{D} -
\bar{D} \bar{\sigma}_a D \right) V_b \, ,\label{Scurrents}
\end{align}
have as lowest $\theta$-components just the current whose name they
bear.  Their respective Ward identities have the form
\begin{align}
\pr^a \hat R_a &= i (D^2 S - \bar D^2 \bar S) + 8 i w \hat\Gamma
\nonumber\\ \pr^a \hat Q_{a \al} &= 8 i w^Q_\al \hat\Gamma \\ \pr^a
\hat T_{ab} &= 8 i w^P_b \hat\Gamma \, , \nonumber
\end{align}
where (see \cite{Piguet})
\begin{align}
w^Q_\al &= i D_\al (D^\be w_\be - \bar{D}_\db \bar{w}^\db ) - 4i
\sigma^a_{\al \da} \pr_a \bar{w}^\da \nonumber\\ w^P_b &= - {\ts
\frac{1}{16}} \bar{\sigma}^{\al \da}_b \left( D^2 \bar{D}_\da w_\al +
\bar{D}^2 D_\al \bar{w}_\da + [D_\al, \bar {D}_\da] (D^\be w_\be -
\bar{D}_\db \bar{w}^\db) \right) \nonumber\\ & \,\, + \half i \pr_b
(D^\be w_\be + \bar D_\be \bar{w}^\db ) \, .
\end{align}

For the $R$ current the breakdown of conformal symmetry manifests
itself in the non-conservation of the current (\ref{sc}) in the
presence of the multiplet $S$, whereas for the supersymmetry current
and the energy momentum tensor it leads to trace contributions of the
form
\begin{align}
\left( \hat Q_a \sigma^a \right)_\da &= 12 i \bar D_\da \bar S + 96 i
\bar{w}_\da \hat\Gamma\, , \nonumber\\ \hat T_a {}^a &= - {\ts
\frac{3}{2}} \left( D^2 S + \bar D^2 \bar S \right) + 12 \, w
\hat\Gamma \, .
\end{align}
The currents corresponding to the remaining symmetries are given by
moments of the $\hat Q$ and $\hat T$ currents \cite{PS}.

\subsection{Wess-Zumino Model on Curved Superspace Background}

For the classical flat space Wess-Zumino model given by the action
\begin{align}\label{WZ}
\hat\Gamma[A,\bar A] &= {\ts \frac{1}{16}} \, \int \! \d^8z \, A \bar
A \, + {\ts \frac{1}{48}} \, g \, \int \! \d^6z \, A^3 \, + \, {\ts
\frac{1}{48}} \, g \, \int \! \d^6\bar{z} \bar A^3 \, \nonumber\\ &
\hspace{3cm} + {\ts \frac{1}{8}} \, m \, \int \! \d^6z \, A^2 \, +
{\ts \frac{1}{8}} \, m \, \int \! \d^6\bar{z} \bar A^2
\end{align}
the Ward identity (\ref{traceidentity}) yields
\begin{eqnarray}
V_{\alpha \da} &=& - {\ts \frac{1}{6}} \left( D_\alpha A \bar D_\da
\bar A - A D_\alpha \bar D_\da \bar A + \bar A \bar D_\da D_\alpha A
\right) \, ,
\label{supercurrent_V}\\
S&=& - {\ts \frac{1}{12}} m A^2 \label{breaking_S}
\end{eqnarray}
for the supercurrent $V_{\alpha \da}$ and the breaking term $S$.  We
observe that the massless Wess-Zumino model is invariant under
superconformal transformations since the breaking term is proportional
to the mass.

For constructing Green functions with multiple insertions of the
supercurrent it is convenient to introduce the external supergravity
field $H^{\al \da}$ conjugate to the supercurrent $V_{\al \da}$,
\begin{gather} \label{VH}
\frac{\delta \Gamma_{\rm ext}}{\delta H^{\al \da}} = {\ts \frac{1}{8}}
V_{\al \da} \, ,
\end{gather}
such that to lowest order in $H^{\al \da}$ we have
\begin{align}
\Gamma &= \hat\Gamma + \Gamma_{\rm ext} = \hat\Gamma + \, {\ts
\frac{1}{8}}\, \int \! \d^8z H^{\al \da} V_{\al \da} \, .
\end{align}
According to the Noether procedure the action $\Gamma$ is given to all
order in $H^{\al \da}$ by the most general diffeomorphism invariant
action which on flat space reduces to (\ref{WZ}). Using the results of
the preceding section we find that this action is given by
\begin{align}  \label{WZc}
\Gamma[A,H^{\al\da}, \phi] &= \Gamma_0[A, H^{\al\da}, \phi] +
\Gamma_S[A, H^{\al\da}, \phi] \\ \Gamma_0[A, H^{\al\da}, \phi] &= {\ts
\frac{1}{16}} \int\!\!\d^8z\,\, E^{-1} A \,\e^{2iH} \bar A \, + \,
{\ts \frac{1}{48}} g\int\!\! \d^6z \,\, \phi^3\,A^3 \, + \, {\ts
\frac{1}{48}} g \int\!\! \d^6\bar z \,\, \bar\phi^3\,\bar A^3
\nonumber\\ \Gamma_S[A, H^{\al\da}, \phi] &= {\ts \frac{1}{8}} m
\int\!\! \d^6z \,\, \phi^3\,A^2 \, + \, {\ts \frac{1}{8}} m \int\!\!
\d^6\bar z \,\, \bar\phi^3\,\bar A^2 + {\ts \frac{1}{8}} \xi \int\!\!
\d^6z \,\, \phi^3\, R\, A^2 \, + \, {\ts \frac{1}{8}} \xi \int\!\!
\d^6\bar z \,\, \bar\phi^3\,\bar R \, \bar A^2 \,. \nonumber
\end{align}
This action is invariant under $\Lambda$ transformations as given in
appendix \ref{transformationlaws}. Infinitesimally this diffeomorphism
invariance is expressed by the equation
\begin{gather} \label{difW}
\left( w_\al\il (H) + w_\al\il(J) + w_\al\il(A) \right) \Gamma = 0 \,
,
\end{gather}
where the operators $w_\al\il(H)$ and $w_\al\il(J)$ are defined by
\begin{gather} \label{superward}
\int \! \d^8z\,\, \delta_\Lambda H^{\al \da} \frac{\delta \Gamma}
{\delta H^{\al \da}} = \int \! \d^8z\,\, \Omega^\al \, w_\al\il(H)
\Gamma \,+c.c. \, , \quad \int \! \d^8z\,\, \delta_\Lambda J \frac{\delta
\Gamma} {\delta J} = \int \! \d^8z\,\, \Omega^\al \, w_\al\il(J)
\Gamma \, ,
\end{gather}
their explicit form being given in appendix \ref{wardoperators} to
second order in $H^{\al \da}$. The Ward identity (\ref{difW}) holds
order by order in the external fields $H^{\al \da}$ and $J$, which can
be checked explicitly.  For this purpose the expansion of $\Gamma$ to
second order in the external fields is given in appendix
\ref{actionexpansion}.

Similarly for the Weyl transformations
\begin{gather}
\int \! \d^6 z \,\, \left( \delta_\sigma A \frac{\delta \Gamma
}{\delta A} \,+ \, \delta_\sigma \phi \frac{\delta \Gamma }{\delta
\phi} \right) = \int \! \d^6 z \,\, \sigma \, w\is(A,J) \Gamma
\end{gather}
yields
\begin{gather}
w\is(A,\phi) \, = \, \phi \frac{\delta}{\delta \phi} \, - \, A
\frac{\delta}{\delta A}
\end{gather}
or similarly
\begin{gather} \label{WeylW}
w\is(A,J) \, = \, \frac{\delta}{\delta J} \, - \, A
\frac{\delta}{\delta A} \, ,
\end{gather}  
which when acting on $\Gamma$ yields the breaking term
\begin{gather} \label{AJswi}
w\is(A,J) \Gamma = - {\ts \frac{3}{2}} S \, ,  \qquad S= - {\ts
 \frac{1}{12}} \left( m \, \e^{3J}A^2 \, - \, \xi \, \e^{3J} R A^2 \,
 + \, \xi \, \e^{3J} \left( \Db^2 + R \right) \e^{2iH}\bar A^2 \,
 \right) \, .
\end{gather}
If $\sigma$ is given in terms of $\Omega_\al$ as in
(\ref{sigmaflach}), we have
\begin{gather}
\int \! \d^6z \, \sigma \, w\is(A,J)\, \Gamma = \, \int \! \d^8z \,
 \Omega^\al w_\al\is(A,J) \, \Gamma
\end{gather}
with
\begin{align}
w_\al\is(A,J) &= \, {\ts \frac{1}{12}} D_\al w\is(A,J) \, ,
\label{walsi}
\end{align}
which yields
\begin{align} \label{swi}
w_\al\is(A,J) \Gamma &= - {\ts \frac{1}{8}} D_\al S \,
\end{align}
for each order in $H^{\al \da}$ and $J$.  Adding (\ref{difW}) and
(\ref{swi}) we find
\begin{gather} w_\al(A,H,J)  \Gamma \equiv
\left( w_\al\il(A,H,J) + w_\al\is(A,J) \right) \Gamma = - {\ts
\frac{1}{8}} D_\al S \label{combiW}
\end{gather}
with $w_\al(A,H,J)$ the Ward operator for the combined chiral
diffeomorphism and Weyl transformations of $H$, $J$ and $A$ whose
explicit forms are given in appendix \ref{wardoperators}.

By inserting the expressions for $\Omega^\al$ given in appendix
\ref{superconformalparameters} into (\ref{Wsconf}), we obtain Ward
operators for the corresponding superconformal transformations,
\begin{equation}
W^X = -i \int\dV \left( \Omega^{X\al} w_\al(A,H,J) \,+\,
\bar\Omega^X_\da \bar w^\da (A,H,J) \right)\,.
\label{Wsconf}
\end{equation}

When restricting to flat space, $H=0$, $J=0$, (\ref{combiW}) yields
again the Ward identity
\begin{align}
-16 w_\al(A) \hat\Gamma &= \bar D^\da V_{\al \da} + 2 D_\al S \,
\end{align}
which agrees with (\ref{traceidentity}) with $V$ and $S$ given by
\begin{align} \label{Vclass}
V_{\alpha \da} &= - {\ts \frac{1}{6}} \left( D_\alpha A \bar D_\da
\bar A - A D_\alpha \bar D_\da \bar A + \bar A \bar D_\da D_\alpha A
\right) -{\ts \frac{1}{3}} \left( \xi \bar {D}_\da D_\al A^2 - \xi
D_\al \bar D_\da \bar A^2 \right) \, , \nonumber\\ S&= - {\ts
\frac{1}{12}} m A^2 + {\ts \frac{1}{12}} \xi \bar D^2 \bar A^2 \, ,
\qquad \bar{S} = - {\ts \frac{1}{12}} m A^2 + {\ts \frac{1}{12}} \xi
D^2 A^2 \, .
\end{align}
Clearly for a conformal theory with $S=\bar S = 0$ we must have $m=0$
and $\xi = 0$.

It should be noted that when imposing the equations of motion
\begin{align}
\frac{\delta  \Gamma}{\delta A} &= 0 \, ,
\end{align}
we have from (\ref{AJswi})
\begin{align}
\frac{\delta  \Gamma}{\delta J} &= - {\ts \frac{3}{2}} S \, ,
\label{onshelltrace}
\end{align}
such that the breaking term $S$ is conjugate to the chiral compensator
$J$ in the same way as the supercurrent $V$ is conjugate to the
superspace metric field $H$ in (\ref{VH}).  The relation
(\ref{onshelltrace}) is frequently used in the literature
\cite{bk},\cite{ggrs}.  However since here we would like to proceed
off-shell we continue to work with the general form (\ref{AJswi}) for
the Weyl transformation.
 
The conformal transformation properties of the supercurrent $V$ may be
obtained from the Ward identity
\begin{gather}
\int \dV \delta_\Lambda H^b \frac{\delta \Gamma}{\delta H^b } + \int\dS
\left(\delta_{\Lambda} A \frac{\delta}{\delta A} + \delta_{\Lambda} J
\frac{\delta}{\delta J} + \delta_\sigma A \frac{\delta}{\delta A} +
\delta_\sigma J \frac{\delta}{\delta J} \right)  \Gamma \, + \,
c.c.  \hspace{4cm} \nonumber\\ \hspace{7.5cm} = \, - {\ts
\frac{1}{12}} \int \dV \Omega^\al D_\al S \, + \, c.c. \, \; .
\end{gather}
Varying this equation with respect to $H^a$ we obtain
\begin{gather} \label{Vconftrans}
\delta V_{\al \da} = - \int \! \d^{\, 8} z \, \Bigg(
\frac{\delta\,(\delta_{\Lambda} H^{\beta \db} |_{O(H)})}{\delta H^{\al
\da}} V_{\beta \db} + \frac{\delta\,(\delta_{\bar \Lambda} H^{\beta
\db} |_{O(H)})}{\delta H^{\al \da}} V_{\beta \db} + \Omega^\beta
D_\beta \frac{\delta S }{\delta H^{\al\da}} + \bar \Omega_\db \bar
D^\db \frac{\delta \bar S }{\delta H^{\al\da}} \Bigg) \, ,
\end{gather}
where the conformal transformation $\delta V_{\al \da}$ is given by
\begin{gather}
\delta V_{\al \da} = \left( \int\dV \delta_\Lambda H^b
 \frac{\delta}{\delta H^b } + \int \dS \left(\delta_{\Lambda} A +
 \delta_\sigma A \right) \frac{\delta}{\delta A} + \int\dS \left(
 \delta _\Lambda J + \delta_\sigma J \right) \frac{\delta}{\delta J}
 +c.c.\right) V_{\al \da}\, .
\end{gather}
The subscript $O(H)$ in (\ref{Vconftrans}) indicates that zeroth order
terms are to be omitted. For $H=J=0$ we find
\begin{align} 
&\int\dS \left(\delta_{\Lambda} A + \delta_\sigma A \right)
\fdq{V_{\al\da}}{A} \,+\,c.c. \nonumber \\ 
&\hspace{1cm}= - \int \! \d^{\, 8}
z \, \frac{\delta\,(\delta_{\Lambda} H^{\beta \db}
|^{(1)}_{O(H)})}{\delta H^{\al \da}} V_{\beta \db} 
- \tfr{3}{2}\int\dS \sigma \fdq{S}{H^{\al\da}}+c.c.
\label{Vconftrans3} \nonumber\\ 
&\hspace{1cm} = \, \quar \left( \left\{ D_\beta, \bar D_\db \right\}
\big( \bar D^\db \Omega^\beta V_{\al \da} \big) + \big(\left\{ D_\al,
\bar D_\da \right\} \bar D^\db \Omega^\beta \big) V_{\beta \db} + D^
\beta \left( \bar D^2 \Omega_\beta V_{\al \da} \right) \right)
\nonumber \\
& \hspace{1cm}\quad \, - \tfr{3}{2}\int\dS \sigma \fdq{S}{H^{\al\da}}
\,+\,c.c. 
\end{align}

By inserting the explicit forms for $\Omega$ given in appendix
\ref{superconformalparameters} we may find the expressions appropriate
for the different conformal transformations.

It may be checked within a BRS type formalism that the representation
for $\delta_\Lambda H$ resp.~$ w\il (H)$ given by (\ref{superward}) is
stable in the sense that any other choice for $\delta_\Lambda H$
satis\-fying the superconformal algebra is equivalent to
(\ref{delta_Lambda_H}) up to redefinitions of $H$.  For this purpose a
constant anticommuting factor is extracted from the $\Lambda$
transformations, such that a corresponding BRS operator $s$ is
obtained,
\begin{equation} \label{slambda}
s \, \Lambda \, = \, \Lambda^2 \, = \, \bigl( \Lambda^{\alpha \da}
D_{\alpha \da} + \Lambda^\alpha D_\alpha + \Lambda_\da \bar D^\da
\bigr) \, \bigl( \Lambda^{\beta \db} D_{\beta \db} + \Lambda^\beta
D_\beta + \Lambda_\db \bar D^\db \bigr) \, ,
\end{equation}
where the $\Lambda$ transformations are now fermionic, such that $s^2
\Lambda =0$ is automatically satisfied.  With $\Lambda$ given by
(\ref{omegadef}) we have, noting that
$\left\{s,D_\alpha\right\}=\left\{s,\bar D^\da \right\}=0$,
\begin{equation} \label{so}
s \Lambda \, = \, - \bar D^\da \, s \Omega^\alpha \, D_{\alpha \da} \,
+\, {\ts \frac{1}{4}} \bar D \bar D \, s \Omega^\alpha \, D_\alpha \,
+ \, {\ts \frac{1}{4}} DD \, s \bar \Omega_\da \bar D^\da \, .
\end{equation}
Inserting (\ref{omegadef}) into (\ref{slambda}) and comparing with
(\ref{so}) implies
\begin{equation} \label{somega}
s \Omega^\alpha = {\ts \frac{1}{2}} \bar D_\db \Omega^\beta \{ \bar
D^\db, D_\beta \} \Omega^\alpha \, + \, {\ts \frac{1}{4}} \bar D \bar
D \Omega^\be D_\be \Omega^\alpha \, + \, \chi^\alpha
\end{equation}
for the superfield $\Omega^\alpha$ which is now bosonic. $\chi^\alpha$
is a chiral field which represents the freedom in expressing $\Lambda$
in terms of $\Omega^\alpha$.  Using this result we may check that
\begin{equation}
s^2 A \, = \, s^2 H^{\alpha \da} \, = \, s^2 J \, = \, 0 \,
\end{equation}   
where $s A,s H^{\alpha \da}, sJ$ are obtained from $\de_\Lambda
 A,\de_\Lambda H^{\alpha \da}, \de_\Lambda J$ given by
 (\ref{delta_Lambda_H}) by inserting (\ref{omegadef}) and by
 extracting a constant anticommuting factor from $\Omega_\alpha$.  By
 complex conjugation a similar result may be obtained for $\bar
 \Omega_\da, \bar A, \bar J$.  To first order in $H$ it is
 straightforward but tedious to check that any supersymmetric and
 Lorentz covariant modification of $sH$ may be absorbed by a suitable
 redefinition of $H$.  For instance modifications of the schematic
 form
\begin{gather}
s' H^{\al \da} = sH^{\al \da} + (\bar D \bar D D \Omega H)^{\al \da}
\, ,
\end{gather}
where all permutations of the symbols in brackets are permitted, are
absorbed by
\begin{gather}
H'{}^{\! \al \da} = H^{\al \da} + (D \bar D H H )^{\al \da} \, ,
\end{gather}
such that
\begin{gather}
 s' H'{}^{\! \al \da} = 0 \, .
\end{gather}

\section{Quantised Theory}

\setcounter{equation}{0}

\subsection{Green Functions}

We quantise the massless Wess-Zumino model by defining its Green
functions by the Gell-Mann-Low formula with suitable subtractions
within the BPHZ scheme. The essence of this renormalization scheme
is to expand integrands whose integrals are potentially divergent
into power series in external momenta, and to subtract terms of
this series such that the integrals over the remaining terms
are well-defined finite expressions. The reader not familiar with
the BPHZ scheme is referred to the literature \cite{bphz}.

For the treatment of massless fields, in which particular care
has to be taken as far as potential IR singularities are concerned, we use
the formalism of Zimmermann and Lowenstein \cite{quantum}. This 
implies that the effective action in the sense of Zimmermann is given
by
\begin{align}\label{WZq}
\hat\Geff [A,\bar A] = &\,{\ts \frac{1}{16}} \hat z\, \int\dV A \bar A
\, + {\ts \frac{1}{48}} \, \hat{g} \, \int \dS A^3 \, + \, {\ts
\frac{1}{48}} \, \hat{g} \, \int \dSb \bar A^3 \, \nonumber\\ & \,-
{\ts \frac{1}{8}} M (s-1) \, \int \dS A^2 \, - {\ts \frac{1}{8}} M
(s-1) \, \int \dSb \bar A^2 \,
\end{align}
on flat space, where $M(s-1)$ is an auxiliary mass parameter with $s$
and $s-1$ being treated like external momenta in the UV and IR
subtractions respectively.  Clearly after performing the subtractions
the massless limit is given by $s=1$.  The Green functions are given
by the perturbative expansion
\begin{gather} \label{Gell-Mann-Low}
\langle A(z_1) \dots A(z_n) \rangle = \, R \, \frac{ \langle
A^{(0)}(z_1) \dots A^{(0) } (z_n) \e^{i \hat\Gamma_{\rm int}(A^{(0)})}
\rangle^{(0)}}{\langle \e^{i \hat\Gamma_{\rm int}(A^{(0)})}
\rangle^{(0)} } \, ,
\end{gather}
where $^{(0)}$ stands for the vacuum expectation value of free fields,
and the interaction is given by
\begin{align}
\hat\Gamma_{\rm int} &= {\ts \frac{1}{16}} \, (\hat z-1) \, \int \dV A
\bar A \, + {\ts \frac{1}{48}} \, \hat{g} \ \, \left( \int \dS A^3 \,
+ \, \int \dSb \bar A^3 \right) \, .
\end{align}
The renormalisation operator $R$ denotes that ultraviolet and infrared
subtractions are performed according to the forest formula with the
subtraction operator
\begin{align} \label{suboperator}
T_\gamma &= \left( 1 - t^{\rho(\gamma) -1}_{p, s-1} \right) \left( 1 -
t^{\delta(\gamma) }_{p, s} \right)
\end{align}
acting on each contribution to the perturbation expansion
corresponding to a 1PI diagram $\gamma$. $\rho(\gamma)$ and
$\delta(\gamma)$ are the infrared and ultraviolet subtraction degrees
associated with this diagram.  $t^{\delta(\gamma) }_{p, s}$ is a
Taylor expansion operator which implies Taylor expansion in the
external momenta $p$ and in the mass parameter $s$ up to and including
order $\delta(\gamma)$.  The subtraction degrees are given by
\begin{align} \label{subdegree}
\delta(\gamma) &= 4 - N_S (1+d_S) - N_V (2+d_V) + \sum\limits{V_i} (
\delta_i - 4) + \half \omega(\gamma) \, , \nonumber\\ \rho(\gamma) &=
4 - N_S (1+d_S) - N_V (2+d_V) + \sum\limits{V_i} ( \rho_i - 4) + \half
\omega(\gamma) \, ,
\end{align}
where $N_S,N_V$ are the number of external (anti-)chiral and vector
fields and $d_S,d_V$ the dimensions of these fields. $\delta_i$ and
$\rho_i$ are the subtraction degrees of the vertex $V_i$ given by the
insertion $\left[ Q^i (z)
\right]^{\rho_i}_{\delta_i}$. $\omega(\gamma)$ is the number of
independent differences $\theta_i - \theta_j , \bar \theta_i - \bar
\theta_j$ in the respective $\theta, \bar \theta$ expansion of the
contribution associated with the graph $\gamma$.  The parameters $\hat
z = 1+ O(\hbar)$ and $\hat g = g+ O(\hbar)$ are fixed by the
normalisation conditions on the vertex functions
\begin{gather} \label{zgcond}
\hat\Gamma_{A \bar A} \Big|_{p^2 = - \mu^2, s=1, \theta=0} =
\tfr{1}{16} \, , \qquad \pr_{\theta_1 }{}^{\!\! 2} \pr_{\theta_2
}{}^{\!\! 2} \hat\Gamma_{A A A} \Big|_{p^2 = q^2 = (p+q)^2 =- \mu^2,
s=1, \theta=0} = {\ts \frac{1}{8}} \, g \, .
\end{gather}
Moreover, as a consequence of the subtraction we have
\begin{gather} \label{mcond}
\pr_{\theta}{}^{\!\! 2} \hat\Gamma_{A A} \Big|_{p=0, s=1} = 0 \, ,
\end{gather}
which guarantees masslessness of the theory.  A fundamental ingredient
for the derivation of quantum Ward identities is the action principle,
which implies for the operator $w_\al(A)$ acting on the vertex
functional $\hat\Gamma$
\begin{align}
-16 \, w_\al(A) \, \hat\Gamma &= -16 \left[w_\al(A)
\,\hat\Geff\right]^{7/2}_{7/2} \cdot \hat\Gamma \, = \left[ \bar D^{\da}
V_{\al \da} + 2 D_\al S \right]^{7/2}_{7/2} \cdot \hat\Gamma \, ,
\end{align}
with
\begin{align}
V_{\al \da} &= - {\ts \frac{1}{6}} \hat z \left( D_\alpha A \bar D_\da
 \bar A - A D_\alpha \bar D_\da \bar A + \bar A \bar D_\da D_\alpha A
 \right) \, , \, \nonumber\\ S &= {\ts \frac{1}{12}} M (s-1) A^2 .
\end{align}

On curved space, the parity invariant effective action is given by
\begin{align} \label{WZcq}
\Geff =\,& \tfr{1}{16}\,\hat z\, \Ikin \,-\,\tfr{1}{8}
\,\left(\Im+\bar\Im\right) \,+\, \tfr{1}{48}\,\hat g\,
\left(\Ig+\bar\Ig\right) \,+\, \tfr{1}{8}\,\hat\xi\,
\left(\Ixi+\bar\Ixi\right) \nonumber\\ &+ \,\tfr{1}{8}\,\hat\lambda_1
\, \left(\Ii+\bar\Ii\right) \,+\, \tfr{1}{8}\,\hat\lambda_2\,
\left(\Iii+\bar\Iii\right)
\end{align}
with
\begin{align}
\Ikin &= \int\dV E^{-1} A \, {\rm e}^{2iH}\bar A & \Ig &= \int\dS
\phi^3A^3 \nonumber\\ \Im &= \int\dS \phi^3\, M(s-1)A^2 & \Ixi &=
\int\dS \phi^3 RA^2 \\ \Ii &= \int\dS \phi^3 R^2 A & \Iii &= \int\dV
E^{-1} A \, {\rm e}^{2iH}\bar R \, .\nonumber
\end{align}
The dynamical fields $A, \bar A$ are quantised, whereas the background
fields $H,J, \bar J$ are treated as classical, i.e. non-propagating.
Possible purely geometrical terms are discarded and will be considered
in a separate publication.

The Green functions are given by the Gell-Mann-Low formula
(\ref{Gell-Mann-Low}) on curved space as well, with the interaction
given by
\begin{align} \label{Gint}
\Gamma_{\rm int} &= \tfr{1}{16} \,(\hat z-1)\, \Ikin \,+\,
 \tfr{1}{48}\,\hat g\, \left(\Ig+\bar\Ig\right) \,+\,
 \tfr{1}{8}\,\hat\xi\, \left(\Ixi+\bar\Ixi\right) \nonumber\\&
 \quad\,+ \,\tfr{1}{8}\,\hat\lambda_1 \, \left(\Ii+\bar\Ii\right)
 \,+\, \tfr{1}{8}\,\hat\lambda_2\, \left(\Iii+\bar\Iii\right) \,.
\end{align}

The external fields yield external, i.e. non-integrated, vertices by
construction.  For the calculation of the subtraction degrees we note
that $H^{\al \da}$ is a vector superfield with $d_{H}= (-1)$ and $J,
\bar J$ are (anti-)chiral superfields with $d_{J}=0$. In the spirit of
perturbation theory all exponentials of the external fields in
(\ref{Gint}) have to be understood as series expansions, such that at
first sight it seems to be non-trivial that the renormalised couplings
$\hat g$ and $\hat \xi$ as well as the field renormalisation $\hat z$
have the same value to all orders in $H,J$ and $\bar J$.  However this
is guaranteed by the requirement of diffeomorphism invariance as
expressed by the Ward identity (\ref{difW}) which we impose also for
the quantum theory.  Our subtraction scheme guarantees that
diffeomorphism invariance is maintained in the quantised theory.
$\hat z (g)$ and $\hat g(g)$ may thus be fixed by the flat space
normalisation conditions (\ref{zgcond}), the masslessness of the
theory still being ensured by (\ref{mcond}). For the renormalised
coupling $\hat\xi(\xi,g)$, we impose the additional vertex function
normalisation condition
\begin{gather} 
\frac{\pr}{\pr p^{H \,a}} \pr_{\theta_2}{}^{\!\! 2}
\pr_{\theta_3}{}^{\!\! 2} \Gamma^a_{H A A }(p_H ; p_2,p_3)
\Big|_{p=p_{sym}, \theta=0, s=1} = \, - {\ts \frac{1}{12}} \xi \,,
\label{xinor}
\end{gather}
where the vertex functions are defined by
\begin{gather} \label{defG}
\Gamma^a_{H A A }(z_H ; z_2,z_3) = \frac{\delta}{\delta H_a(z_H)}
\frac{\delta}{\delta A(z_2)} \frac{\delta}{\delta A(z_3)} \Gamma
\Big|_{A=H=J=0}
\end{gather}
In (\ref{defG}) all coordinates are superspace coordinates.
Insertions of the supercurrent and of the trace are given by
\begin{align}
\frac{\delta}{\delta H_a(z)} \Gamma &=
\left[\fdq{\Geff}{H_a(z)}\right]_3^3 \cdot\Gamma = \left[V^a[A,\bar
A,H,J,\bar J](z) \right]^3_3 \cdot \Gamma \, , \nonumber\\ w\is(A,J)
\Gamma &= \left[ w\is(A,J) \Geff\right]_3^3 \cdot \Gamma \equiv - {\ts
\frac{3}{2}} \left[S [A,\bar A,H,J,\bar J] (z) \right]^3_3 \cdot
\Gamma \, , \nonumber\\ w^{\! (\bar \sigma)}(\bar{A}, \bar{J}) \Gamma
&= \left[ w^{(\bar\sigma)}(\bar A, \bar J) \Geff \right]_3^3
\cdot\Gamma \equiv - {\ts \frac{3}{2}} \left[\bar{S} [A,\bar
A,H,J,\bar J](z)\right]^3_3 \cdot \Gamma \, ,
\label{insertionsdef}
\end{align}
from which 1PI Green functions with insertion may be obtained by
virtue of
\begin{gather}
\frac{\delta}{\delta A(z_n)} \dots \frac{\delta}{\delta A(z_1)}
\frac{\delta}{\delta H_a(z)} \Gamma \Big|_{H=J=\bar J=0,A=\bar
A=0} = \langle \left[V^a[A,\bar A](z) \right]^3_3 A(z_1) \dots A(z_n)
\rangle^{1PI} \, ,
\end{gather}
and similarly for the traces.  The trace insertions $[S]^3_3$ and
$[\bar S]^3_3$ contain oversubtracted terms of the form \mbox{$[M
(s-1) \phi^3 A^2]^3_3$} and $[M (s-1) \bar \phi^3 \bar A^2]^3_3$,
where the expression $(s-1)$ has to be treated like an external
momentum in the subtractions, such that some additional information
has to be used before the massless limit $s=1$ may be taken.  This is
provided by the Zimmermann identities which express oversubtracted
insertions by the corresponding mini\-mally subtracted ones plus a
basis of local field polynomial insertions with the original
subtraction degree \cite{zi}. For our purposes it is convenient to use
the Zimmermann identities in their non-integrated form.  For the
chiral and antichiral mass terms we obtain
\begin{align} \label{chZ}
\left[ M (s-1) \phi^3 A^2 \right]^3_3 \cdot \Gamma \Big|_{s=1}& =
 \Big[ \ukin \phi^3 A \left(\Db^2 + R \right) \bar {A} + \ug \phi^3
 A^3 \nonumber\\ & \;\;\;\; + \uxi \phi^3 R A^2 + \uxib \phi^3 \left(
 \Db^2 + R \right) \bar {A}^2 \nonumber\\ & \;\;\;\; + \ui \phi^3 R^2
 A +\uib \phi^3 \left(\Db^2+R\right) (\bar A\bar R) \nonumber\\ &
 \;\;\;\;+\uii \phi^3 \left( \Db^2 + R\right) \left( \D^2 + \bar
 R\right) A \nonumber \\ & \;\;\;\; +\uiib \phi^3 R \left( \Db^2 +
 R\right) \bar A +\ubox \phi^3 \Box A \Big]^3_3 \cdot \Gamma
 \Big|_{s=1} \,\, , \\
\left[ M (s-1) \bar\phi^3 \bar A^2 \right]^3_3 \cdot \Gamma
 \Big|_{s=1}& = \Big[ \ukin \phi^3 \bar A \left( {\D}^2 + \bar R
 \right) {A} + \ug \bar\phi^3 \bar A^3 \nonumber\\ & \;\;\;\; + \uxi
 \bar \phi^3 \bar R \bar A^2 + \uxib \bar \phi^3 \left( {\D}^2 + \bar
 R \right) {A}^2 \nonumber\\ & \;\;\;\; + \ui \bar \phi^3 \bar R^2
 \bar A +\uib \bar \phi^3 \left(\D^2+\bar R\right) ( A R) \nonumber\\
 & \;\;\;\; +\uii \bar\phi^3 \left( {\D}^2 + \bar R\right)\left( \Db^2
 + R\right)\bar A \nonumber \\ & \;\;\;\; +\uiib \bar \phi^3 \bar R
 \left( {\D}^2 + \bar R\right) A +\ubox \bar \phi^3
 \bar{\rule{0cm}{1.7ex}\Box}\bar A \Big]^3_3 \cdot \Gamma \Big|_{s=1}
 \,\, ,
\label{achZ} 
\end{align}
where the chiral d'Alembertian is given by
\begin{gather}
\Box \, A = \left( \Db^2+R \right) \D^2 \, A \, , \label{defbox}
\end{gather}  
and, again, purely geometrical terms have been omitted.  In
(\ref{chZ})--(\ref{defbox}) all representation changing factors
$\e^{\pm 2iH}$ have been suppressed.  The Zimmermann identities
contain all (anti-)chiral local field polynomials which are
diffeomorphism invariant and of dimension $3$.  Parity invariance
requires all coefficients $u$ to be real.

\subsection{Callan-Symanzik equation}
Conformal transformations are in general anomalous to higher orders.
On the level of integrated Ward identities, these anomalies are
parametrised by the Callan-Symanzik (CS) functions $\beta$ and
$\gamma$.  These functions have to be identified before proceeding to
the investigation of local Ward identities.

\subsubsection{Independent couplings}
As a first application of the Zimmermann identities
(\ref{chZ}),(\ref{achZ}) we derive a CS equation in which the coupling
$\xi$ of $A^2$ to the curvature $R$ is taken to be independent of the
elementary coupling $g$. In this case the counterterm coefficient
$\hat\xi=\hat\xi(\xi,g)$ is fixed by the normalisation condition
(\ref{xinor}).  Similarly, the coefficients of the linear terms in
$\Geff$ are treated as independent and have to be fixed by some
normalisation conditions.  We claim that a CS equation
\begin{gather}
\left.\C\Gamma\right|_{s=1} = 0 \label{CSeq}\\[-0.2cm]
\intertext{holds with} \C=m\partial_m + \beta_g \partial_g + \beta_\xi
\partial_\xi + \beta_{\lambda_1} \partial_{\lambda_1} +
\beta_{\lambda_2} \partial_{\lambda_2} -\gamma \N\,, \label{CSop}\\
\intertext{where $m\partial_m$ includes all mass parameters of the
theory and} \N=\int \dS A \fdq{}{A} + \int\dSb \bar A \fdq{}{\bar A}
\end{gather}
is the counting operator for matter legs. For the proof we apply $\C$
to $\Gamma$, use the action principle and the Zimmermann identity and
find that
\begin{align}
\left.\C\Gamma\right|_{s=1} = \Big[ & \phantom{+}
  \,\,{\ts\frac{1}{16}} \Ikin && \Big( \beta_g\partial_g z - 2\gamma
  \hat z -4(1-2\gamma)\ukin \Big) \nonumber\\ &+{\ts\frac{1}{48}}
  (\Ig+\bar\Ig) && \Big( \beta_g\partial_g \hat g - 3\gamma \hat g
  \Big)\nonumber\\ &+ {\ts\frac{1}{8}} (\Ixi+\bar\Ixi) && \Big(
  (\beta_g\partial_g + \beta_\xi\partial_\xi +
  \beta_{\lambda_1}\partial_{\lambda_1} +
  \beta_{\lambda_2}\partial_{\lambda_2}) \hat \xi \nonumber\\
  &\phantom{+++} && - 2\gamma\hat\xi- (1-2\gamma) (\uxi + \uxib)
  \Big)\, \nonumber\\ &+ {\ts\frac{1}{8}} (\Ii+\bar\Ii) && \Big(
  (\beta_g\partial_g + \beta_\xi\partial_\xi +
  \beta_{\lambda_1}\partial_{\lambda_1} +
  \beta_{\lambda_2}\partial_{\lambda_2}) \hat \lambda_1 \\
  &\phantom{+++}&& -\gamma \hat \lambda_1 - (1-2\gamma) (\ui + \uib)
  \Big)\, \nonumber\\ &+ {\ts\frac{1}{8}} (\Iii+\bar\Iii) && \Big(
  (\beta_g\partial_g + \beta_\xi\partial_\xi +
  \beta_{\lambda_1}\partial_{\lambda_1} +
  \beta_{\lambda_2}\partial_{\lambda_2}) \hat \lambda_2 \nonumber\\
  &\phantom{+++}&& -\gamma \hat \lambda_2- (1-2\gamma) (\uii + \uiib)
  \Big)\, \quad & \Big]_4^4 \,\cdot \, \Gamma \, \Big|_{s=1}
  \,.\hspace*{2cm}\nonumber
\end{align}
Hence (\ref{CSeq}) is true if the system
\begin{align}
\beta_g\partial_g \hat z - 2\gamma \hat z -4(1-2\gamma)\ukin\, &=0
\nonumber\\ \beta_g\partial_g \hat g - 3\gamma \hat g\, &=0
\nonumber\\ ( \beta_g\partial_g + \beta_\xi\partial_\xi +
\beta_{\lambda_1}\partial_{\lambda_1} +
\beta_{\lambda_2}\partial_{\lambda_2} - 2\gamma )\hat\xi - (1-2\gamma)
(\uxi + \uxib) \,&=0 \\ (\beta_g\partial_g + \beta_\xi\partial_\xi +
\beta_{\lambda_1}\partial_{\lambda_1} +
\beta_{\lambda_2}\partial_{\lambda_2} -\gamma) \hat \lambda_1 -
(1-2\gamma) (\ui + \uib) \,&=0 \nonumber \\ (\beta_g\partial_g +
\beta_\xi\partial_\xi + \beta_{\lambda_1}\partial_{\lambda_1} +
\beta_{\lambda_2}\partial_{\lambda_2} -\gamma) \hat \lambda_2 -
(1-2\gamma) (\uii + \uiib) \, &=0 \nonumber
\end{align}
is satisfied.  This system has a unique solution for $\beta$ and
$\gamma$ order by order in perturbation theory, hence (\ref{CSeq}) is
proved. It may be shown in general that both the vertex functional $\Gamma
$ as well as the $\beta$ and $\gamma$ functions are independent of
the auxiliary mass $M$ \cite{PS}.

\subsubsection{Dependent couplings and $R$ invariance}
\label{reductionsection}
Inspired by analogous results in $\phi^4$ theory where the improvement
coefficient is a unique function of the elementary coupling
\cite{KS2}, we wish to understand $\xi$, $\lambda_1$ and $\lambda_2$
as functions of $g$ in the supersymmetric case too.  Necessary and
sufficient conditions for such a functional dependence to be
consistent with renormalisability are the reduction equations \cite{Reduction}
\begin{equation}
\beta_g \frac{{\rm d}p}{{\rm d} g} = \beta_p , \qquad
  p=\xi,\lambda_1,\lambda_2, \label{redeqp} \qquad p = p(g) \,.
\end{equation}
These imply that the equations
\begin{align}
\beta_g\partial_g \hat\xi &= 2\gamma \hat \xi + (1-2\gamma)
(\uxi+\uxib)\nonumber\\ \beta_g\partial_g \hat\lambda_1 &= \gamma \hat
\lambda_1 + (1-2\gamma) (\ui+\uib) \label{redeq} \\ \beta_g\partial_g
\hat\lambda_2 &= \gamma \hat \lambda_2 + (1-2\gamma)
(\uii+\uiib)\nonumber
\end{align}
have to be satisfied by $\hat\xi$, $\hat\lambda_{1,2}$.  In order to
show that this system of equations has a solution we proceed as
follows.  We show that there exists a particular choice for $\hat\xi$,
$\hat\lambda_{1,2}$ such that the corresponding theory is $R$
invariant.  Consistency of the $R$ transformation and Callan Symanzik
operators implies for this particular choice that the reduction
equations (\ref{redeq}) are satisfied, which in turn implies
renormalisability of the theory.

As mentioned above (\ref{Vclass}) the classical theory is $R$
invariant for
\[\xi=\hat\xi^{(0)} =0 \qquad \lambda_{1,2} =
\hat\lambda_{1,2}^{(0)}=0\,. \] To higher orders, $R$ invariance can
be maintained by setting
\begin{align}
\hat\xi &= \half ( \uxib-\uxi) \nonumber \\ \hat\lambda_1 &= \quar (
\uib-\ui ) \label{Rinveq}\\ \hat\lambda_2 &= \half ( \uii-\uiib
)\,. \nonumber
\end{align}
By studying consistency of the CS operator with $R$ invariance we now
show that the reduction solution coincides with the $R$ invariant
theory, such that reduction implies a symmetry (and vice versa).  The
Ward operator for $R$ transformation is given by (\ref{Wsconf}) with
$\Omega^{R\al}$ from appendix \ref{superconformalparameters}, and we
have
\[\left.W^R \Gamma\right|_{s=1} = 0.\]
Furthermore the consistency condition
\[
[\C,W^R] =0
\]
holds with $\C = m\partial_m + \beta_g \partial_g -\gamma\N$.  As a
consequence
\begin{equation}
\left.W^R \C \Gamma \right|_{s=1} =0 \,.\label{RCG}
\end{equation}

We now proceed by induction. At order $\hbar^0$ the system
(\ref{redeq}) is trivial. Assuming that it holds for orders up to and
including $n$ we have for $\C\Gamma$:
\begin{align}
\left.\C\Gamma\right|_{s=1} = \phantom{+} &\, {\ts\frac{1}{8}}
   \Big(\left( \beta_g\partial_g - 2\gamma \right) \hat\xi -
   (1-2\gamma) (\uxi + \uxib) \Big)\, (\Ixi+\bar\Ixi) \nonumber\\ +&\,
   {\ts\frac{1}{8}} \Big( \left(\beta_g\partial_g -\gamma \right) \hat
   \lambda_1 - (1-2\gamma) (\ui + \uib) \Big)\, (\Ii+\bar\Ii)
   \nonumber\\ +&\, {\ts\frac{1}{8}} \Big( \left( \beta_g\partial_g
   -\gamma \right) \hat \lambda_2 - (1-2\gamma) (\uii+\uiib) \Big)\,
   (\Iii+\bar\Iii) \nonumber \\ +&\, o(\hbar^{n+2}) \,.
\end{align}
The l.h.s. is of order $\hbar^{n+1}$ by assumption.  To this order the
r.h.s. is local since the nonlocal contributions are of order $n+2$.
Applying $W^R$ and using (\ref{RCG}) we have to order $n+1$
\begin{align}
0 = \phantom{+} &\,{\ts\frac{1}{8}} \Big(\left( \beta_g\partial_g -
2\gamma \right) \hat\xi - (1-2\gamma) (\uxi + \uxib) \Big)\,\cdot\,
\tfr{4}{3} ( \Ixi- \bar\Ixi) \nonumber\\ +&\, {\ts\frac{1}{8}} \Big(
\left(\beta_g\partial_g -\gamma \right) \hat \lambda_1 - (1-2\gamma)
(\ui+\uib) \Big)\, \cdot\, \tfr{8}{3} ( \Ii- \bar\Ii) \nonumber\\ +&\,
{\ts\frac{1}{8}} \Big( \left( \beta_g\partial_g -\gamma \right) \hat
\lambda_2 - (1-2\gamma) (\uii + \uiib) \Big)\, \cdot \, \tfr{4}{3}
(\bar \Iii- \Iii)\, .
\end{align}
Hence (\ref{redeq}) holds to order $n+1$, which completes the
proof. Thus the quantum theory with $g$ as the only independent
coupling is well defined and renormalisable, and $R$ invariance is
realised at $s=1$.

As a check we calculate $\beta$ to one-loop in appendix
\ref{betafunction} and find
\begin{align}
\beta^\xi &= \left(a_{10} + a_{11} \xi \right) g^2 + \left(a_{20} +
a_{21} \xi \right) g^4 + \left(a_{30} + a_{31} \xi \right) g^6 + \dots
\, \\ \beta_\xi^{(0)} &= 2\gamma^{(1)}\xi \,,\nonumber
\end{align}
with $a_{10}=0$, $a_{11}=\frac{1}{(4\pi)^2}$.  With
\begin{equation}
\xi(g)= \xi_0 + \xi_1 g^2 + \xi_2 g^4 + \dots
\end{equation}
the reduction equation (\ref{redeqp}) yields \\[0.5cm]
\begin{tabular}{rll}
at & $o(\hbar)$: & $ a_{10} + a_{11} \xi_0 = 0$ \\ & $o(\hbar^2)$: & $
   \xi_1 (3 {\ts \frac{1}{(4 \pi)^2}} - a_{11}) = a_{20} + a_{21}
   \xi_0 $\\ & $o(\hbar^{n+1})$: & $ \xi_n ( 3 n {\ts \frac{1}{(4
   \pi)^2}} - a_{11}) =f_n(\xi_i, i < n) $ \,.
\end{tabular}\\[0.5cm]
Hence $\xi_0=0$ and all $\xi_n$ are uniquely determined in accordance
with the general arguments above.  This completes the analysis of
dilatational anomalies at the level of integrated Ward identities.

\subsubsection{Local Callan-Symanzik equation}
We proceed to consider local Ward identities in order to analyse the
anomalies of the full superconformal group, which is necessary in view
of determining the current algebra.  For this purpose we introduce a
local $[\dots]_3^3$-insertion $\Leff$ whose integral yields $\Geff$,
\begin{align}
\Geff &= \int\dS \Leff + \int\dSb \Leffb \,,\\[0.2cm] 
\Leff = &
\phantom{+} \tfr{1}{32}\hat z \Lkin + \tfr{1}{48} \hat g \Lg
-\tfr{1}{8} \Lm \nonumber\\ & + \tfr{1}{8} \,\,(\hat\xi
-\hat\eps) \,\,\Lxi \,+ \tfr{1}{8} \hat\eps \,\,\Lxib
\rule{0cm}{3ex}\nonumber\\ & + \tfr{1}{8} (\hat\lambda_1 -\hat\eta_1)
\Li + \tfr{1}{8} \hat\eta_1 \Lib \rule{0cm}{3ex}\\ & + \tfr{1}{8}
(\hat\lambda_2-\hat\eta_2)\Lii + \tfr{1}{8} \hat\eta_2 \Liib
\rule{0cm}{3ex}\nonumber\\ & + \tfr{1}{8} \hat\alpha \Lbox\, .
\rule{0cm}{3ex}\nonumber
\end{align}
Here we have introduced
\begin{equation}
\begin{array}{rclcrcl}
\Lkin &=& \phi^3 A \left(\Db^2 + R \right) \e^{2iH}\bar {A} & & \Lg
&=&\phi^3 A^3 \\ \Lm &=& \phi^3M(s-1)\phi^3 A^2 & & \Lbox&=&
\phi^3\left( \Db^2+R \right) \e^{2iH}\D^2\e^{-2iH} \, A
\rule{0cm}{3ex}\\ \Lxi &=& \phi^3 RA^2 & & \Lxib &=& \phi^3 \left(
\Db^2 + R \right) \e^{2iH}\bar {A}^2 \rule{0cm}{3ex}\\ \Li &=& \phi^3
R^2 A & & \Lib &=& \phi^3 \left( \Db^2 + R \right) \e^{2iH}(\bar A\bar
R) \rule{0cm}{3ex}\\ \Lii &=& \phi^3 \left( \Db^2 + R \right)
\e^{2iH}(\D^2+\bar R) \e^{-2iH}A & & \Liib &=& \phi^3 R \left( \Db^2 +
R \right) \e^{2iH}\bar A \,,\rule{0cm}{3ex}
\end{array} \label{Ldef}
\end{equation}
such that
\begin{align}
\int\dS\Lkin&= \Ikin & \int\dS\L_{g,M} &= I_{g,M} & \int\dS\Lbox &=0
\nonumber\\ \int\dS\Lxi &= \Ixi & \int\dS\Li &= \Ii & \int\dS\Lii &=
\Iii \\ \int\dS \Lxib &= \bar\Ixi & \int\dS \Lib &= \bar\Ii &
\int\dS\Liib &= \bar\Iii \,. \nonumber
\end{align}
The functions $\hat\eps=\hat\eps(g)$,
$\hat\eta_{1,2}=\hat\eta_{1,2}(g)$ and $\hat\alpha=\hat\alpha(g)$
represent ambiguities in the choice of the local basis for $\Leff$ and
the corresponding terms cancel when integrated.

We aim at proving an equation which expresses broken Weyl invariance
at the local level, of the form
\begin{equation}
\left( w^{(\sigma)} - \gamma A\fdq{}{A} \right) \Gamma\Big|_{s=1} +
 \left[ \beta_g \partial_g \Leff \right] \,\cdot\, \Gamma \Big|_{s=1}
 =0\,,
\label{localCS}
\end{equation}
as well as a corresponding antichiral equation.  We denote these
equations by {\it local CS equations}.

Analogously to the integrated case we find that (\ref{localCS}) is
satisfied if the following equations hold, \newlength{\mylength}
\settowidth{\mylength}{$\beta_g\partial_g (\hat\lambda_1-\hat\eta_1) -
(1-2\gamma) \ui -(2+\gamma)\hat\lambda_1$}
\begin{subequations}
\begin{align}
\beta_g\partial_g \hat z - 2\gamma \hat z -4(1-2\gamma)\ukin\, &=0
\label{loceq1a} \\ \makebox[\mylength][r]{$ \beta_g\partial_g \hat g -
3\gamma \hat g$} &=0 \label{loceq1b}
\end{align}
\end{subequations}
\vspace*{-1cm}
\begin{subequations}
\begin{align}
\beta_g\partial_g (\hat\xi-\hat\eps) - (1-2\gamma) \uxi
-(1+2\gamma)\hat\xi &=0 \label{loceq2a}\\
\makebox[\mylength][r]{$\beta_g\partial_g \hat\eps - (1-2\gamma) \uxib
+\hat\xi$} &=0 \label{loceq2b}
\end{align}
\end{subequations}
\vspace*{-1cm}
\begin{subequations}
\begin{align}
\beta_g\partial_g (\hat\lambda_1-\hat\eta_1) - (1-2\gamma) \ui
-(2+\gamma)\hat\lambda_1 &=0 \label{loceq3a}\\
\makebox[\mylength][r]{$\beta_g\partial_g \hat\eta_1 - (1-2\gamma)
\uib +2\hat\lambda_1$} &=0 \label{loceq3b}
\end{align}
\end{subequations}
\vspace*{-1cm}
\begin{subequations}
\begin{align}
\beta_g\partial_g (\hat\lambda_2-\hat\eta_2) - (1-2\gamma) \uii
+(1-\gamma)\hat\lambda_2 &=0 \label{loceq4a}\\
\makebox[\mylength][r]{$\beta_g\partial_g \hat\eta_2 - (1-2\gamma)
\uiib -\hat\lambda_2$} &=0 \label{loceq4b}
\end{align}
\end{subequations}
\vspace*{-1cm}
\begin{align}
\makebox[\mylength][r]{$\beta_g\partial_g \hat\alpha
-(1-2\gamma)\ubox$} &=0 \,.
\label{loceq5}
\end{align}
Equations (\ref{loceq1a}),(\ref{loceq1b}) are the usual equations for
$\beta_g$ and $\gamma$, and hence are satisfied.\\ The sums
(\ref{loceq2a})+(\ref{loceq2b}), (\ref{loceq3a})+(\ref{loceq3b}) and
(\ref{loceq4a})+(\ref{loceq4b}) constitute the reduction equations
(\ref{redeq}) for $\hat\xi$, $\hat\lambda_1$ and $\hat\lambda_2$,
which are satisfied by the theory constructed in section
\ref{reductionsection}.  Inserting these into (\ref{loceq2b}),
(\ref{loceq3b}) and (\ref{loceq4b}) and using (\ref{Rinveq}), which
also holds in this theory, yields
\begin{subequations}
\begin{align}
\beta_g\partial_g \hat\eps &= \half \beta_g\partial_g\hat\xi
-3\gamma\hat\xi
\label{loceq2c}\\
\beta_g\partial_g \hat\eta_1 &= \half \beta_g\partial_g\hat\lambda_1
-\tfr{9}{2}\gamma\hat\lambda_1\label{loceq3c}\\ \beta_g\partial_g
\hat\eta_2 &= \half \beta_g\partial_g \hat\lambda_2 +\tfr{5}{2}
\hat\lambda_2 \,. \label{loceq4c}
\end{align}
\end{subequations}
For establishing the local CS equation (\ref{localCS}), we have to
provide $\hat\eps$, $\hat\eta_1$, $\hat\eta_2$, $\hat\alpha$ such that
equations (\ref{loceq2c}), (\ref{loceq3c}), (\ref{loceq4c}) and
(\ref{loceq5}) are satisfied.

Inserting the power series expansions
\begin{align}
\beta_g &= \sum_{n=1}^{\infty} \beta_n \hbar^n g^{2n+1}\,, & \gamma &=
\sum_{n=1}^{\infty} \gamma_n \hbar^n g^{2n}\,\\ \intertext{and}
\hat\eps &= \sum_{n=0}^{\infty} \hbar^n \hat\eps_n(g), & \hat\xi &=
\sum_{n=0}^{\infty} \hbar^n \hat\xi_n(g)
\end{align}
into (\ref{loceq2c}), we have to order $N$ in $\hbar$
\begin{align}
\beta_1 g^3 \partial_g \hat\eps_{N-1}(g) &= \sum_{k=1}^{N} q_k(g) -
\sum_{k=1}^{N} \beta_k g^{2k+1} \partial_g \hat\eps_{N-k}(g)\, , \\
q_k(g) &= \half \beta_k g^{2k+1} \partial_g \hat\xi_{N-k}(g) -3
\gamma_k g^{2k} \hat\xi_{N-k}(g)\, . \nonumber
\end{align}
We assume that the coefficients $\hat\eps_1,\dots,\hat\eps_{N-2}$ have
already been determined. Then, since $\be_1=\frac{3}{2(4\pi)^2}\neq 0$
and the r.h.s. is a power series starting at least with $g^3$, the
coefficient $\hat\eps_{N-1}$ can be found by trivial integration, such
that $\hat\eps_n$ is a {\it power series} in $g$, once we choose the
arbitrary integration constant to be $\hat\eps_0(g)=0$.  The same
argument applies to the other differential equations. Hence we have
established the local CS equation (\ref{localCS}).

\section{Transformation properties}
\label{sectransprop}

\setcounter{equation}{0}

In this section we calculate the superconformal transformation
properties of the supercurrent in the quantum theory using the local
Ward identities (\ref{combiW}) and (\ref{localCS}) as a starting
point.  First let us recall that the supercurrent is defined to all
orders in (\ref{insertionsdef}) with $\hat\xi$, $\hat\lambda_1$,
$\hat\lambda_2$ fixed by (\ref{Rinveq}). The trace $S$, originally
given by (\ref{insertionsdef}), has been brought into the form
(\ref{localCS}), where the symmetry breaking term is separated into
anomalous contact terms of the quantised fields $A$ and into an
operator insertion such that the character of an $S$ type conformal
breaking is maintained.

Since the quantised theory is diffeomorphism invariant by
construction, the symmetry breaking which we look for is entirely
determined by the breaking of Weyl symmetry in curved space.  In
particular it has precisely the form of the tree approximation, hence
(\ref{combiW}) may be rewritten to all orders in terms of insertions,
\begin{equation}
\int\dV \Omega^\al w_\al^{(\gamma)}(A,J)\,\Gamma + c.c. = - \int\dV
\delta_\Omega H^{\al\da} \fdq{\Gamma}{H^{\al\da}} -\tfr{3}{2}\int\dS
\sigma S\cdot\Gamma +c.c. \,,
\label{quantWI}
\end{equation}
where
\begin{align}
w_\al^{(\gamma)} (A,J) &\equiv w_\al\il(A,J)+ w_\al\is(A,J) -\tfr{1}{12} \gamma D_\al
\left( A \fdq{}{A}\right)\,,\\ S &\equiv \tfr{2}{3}\left[ \beta_g\partial_g
\Leff \right]_3^3 \, ,
\end{align}
and $w_\al\is(A,J)$ is given by (\ref{walsi}).

Differentiation of (\ref{quantWI}) with respect to $H$ yields the
transformation properties of Green functions with supercurrent
insertions as required.  For the first derivative we obtain, when
restricting to flat space,
\begin{align}
\lefteqn{ \int\dV \Omega^\al w_\al^{(\gamma)}(A)
[V_{\al\da}(z')]\cdot\hat\Gamma + c.c.} &\nonumber\\ &\qquad\qquad
=-\int\dV \fdq{(\delta_\Omega H^{\be\db}{}^{(1)})}{H^{\al\da}(z')}
[V_{\be\db}] \cdot\hat\Gamma - 12\left(\fdq{}{H^{\al\da}(z')}\int\dS
\sigma S\cdot\Gamma\right)\Big|_{H=0} +c.c.  \nonumber\\ &\qquad\qquad
= [\delta^{({\rm cl})} V_{\al\da}(z')]\cdot\hat\Gamma - \Bigg\{ 8\int\dS
\sigma \left[\beta_g\partial_g\fdq{\Leff}{H^{\al\da}(z')}
\Big|_{H=0}\right]\cdot \hat\Gamma \nonumber\\
&\qquad\qquad\qquad\qquad\qquad\qquad + \int\dS \sigma
[\beta_g\partial_g\hat\Leff]\cdot[V_{\al\da}(z')]\cdot\hat\Gamma
+c.c.\Bigg\} \,.
\label{Vquanttrans}
\end{align}
Only the linear part of $\delta_\Omega H$ contributes since the terms
involving the inhomogenous parts cancel for superconformal
transformations which satisfy $\bar
D^\da\Omega^\al=D^\al\bar\Omega^\da$ by definition. The hats $\hat{}$
denote that for the corresponding quantities $H=J=0$. Equation
(\ref{Vquanttrans}) holds to all orders and reduces to
(\ref{Vconftrans}) in the classical approximation, where $S$ vanishes
and all contributions are purely local.

The transformation laws for vertex functions with multiple insertions
of the supercurrent can be obtained by multiple differentiation of
(\ref{quantWI}) with respect to $H$.  Such differentiations never
involve the mass term hence the massless limit $s=1$ may be taken
without the necessity of using a Zimmermann identity.  This means that
(\ref{quantWI}) contains all information about the transformation
properties of the supercurrent even when multiply inserted into vertex
functions.  $S$ type breaking of superconformal symmetry ensures that
the multiple insertions transform covariantly under translations,
Lorentz and supersymmetry transformations.  All other transformations
($R$, dilatations, special conformal, special supersymmetry) lead to
anomalies.  However there are no further anomalies than those already
present for Green functions of elementary fields.  This is the main
result of our analysis.

Supercurrent and trace are in direct relation with those of equation
(16.1.15) of \cite{PS} for $s=1$ (S type) in the flat space limit, if
we define $L=\frac{\hat z}{32} \bar D^2 (A\bar A) -\frac{1}{8} M(s-1)
A^2 +\frac{\hat g}{48} A^3$, such that on flat space $\Leff$ given by
(\ref{localCS}) and $L$ differ by the terms involving $\hat\xi$,
$\hat\epsilon$, $\hat\lambda_1$, $\hat\lambda_2$, $\hat\eta_1$,
$\hat\eta_2$, $\hat\alpha$. In \cite{PS}, the trace is taken to be $S'
= \tfr{2}{3} \beta_g\partial_g L$ and the coefficients of the terms
present in $\Leff-L$ differ accordingly in the supercurrent when
compared to the supercurrent discussed in this paper.  Using
(\ref{loceq2a}) -- (\ref{loceq5}) of the present paper we can show
that the difference $V'_{\al\da} - V_{\al\da}$ is in exact agreement
with $S-S'$ such that in both cases
\[ \bar D^\da [V_{\al\da}]\cdot\Gamma = -16w_\al^{(\gamma)} \Gamma -2
D_\al [S]\cdot\Gamma \] is satisfied.  This demonstrates that there is
an arbitrariness in the definition of $V_{\al\da}$ and $S$ on flat
space. However the definitions of this paper have the feature of
coinciding with the well-defined flat space limit of canonical
supergravity.

\section{Discussion and conclusions}
\label{secdisc}

\setcounter{equation}{0}

In the present paper we have derived the superconformal transformation
properties of Green functions with multiple insertions of the
supercurrent.  We have worked to all orders of perturbation theory
within the context of the massless Wess-Zumino model. The main tool of
our analysis is the embedding of the model into curved superspace with
the usual supergravity constraints and a chiral
compensator. Supersymmetry and diffeomorphism invariance can be kept
manifest, such that all superconformal transformations can be
expressed by Weyl transformations. This applies also to higher orders
and permits the exhaustive study of possible anomalies. Since our
analysis employs Ward identities, it is scheme independent.

As a conclusion we compare our results with the literature.  Close in
method and spirit is the analogous investigation
\cite{KS3}--\cite{KS2} in $\phi^4$ theory. The most remarkable
difference between the two models is the fact that the $S$ type
supercurrent contains an {\it improved} energy-momentum tensor.  While
in $\phi^4$ theory the improvement has to be built in and necessitates
some technically involved steps in the subsequent analysis, it is
automatic in the supersymmetric case.  In $\phi^4$ theory it is not
possible to separate trace effects and hence to define multiple
insertions with controlled transformation behaviour easily
(s. \cite{KS2} sect. 5), even when the couplings are generalised to be
local fields.  Here there is no difficulty in doing so even without
local couplings.  This is an important feature which indicates that
further structural relations may also be derived more easily in the
supersymmetric case than in the non-supersymmetic one.

Complementary to our approach is the investigation performed in
\cite{PW}.  There the general aim is to quantise supersymmetric
theories in terms of conventional fields and thus to have a simple
transition to conventional non-supersymmetric field theories.  This
effort is essential if one aims at having contact with phenomenology.
Supersymmetry transformations become non-linear and cannot be realised
trivially.  Supermultiplets have a non-trivial form too.  This makes a
literal translation between the two approaches difficult.  In
particular it is not obvious how the two supergravity multiplets are
related.  In any case, comparison in detail is somewhat premature.
Whereas in \cite{PW} a general supersymmetric gauge theory is
considered and here only a non-gauge model is studied, the present
analysis treats already higher orders.  We hope to come back to this
comparison in the near future.

In order to give an outlook to applications and further investigations
we complete the Weyl identity, i.e. the local CS equation, by adding
purely geometrical terms $ C(H,J,\bar J)$
\begin{equation}
\left(w\is-\gamma A\fdq{}{A}\right)\Gamma - \beta_g
[\partial_g\Leff]\cdot\Gamma = C(H,J,\bar J)\,
\label{geomterms}.
\end{equation}
These have to be Weyl invariant and contain at least the
supersymmetric version of the Weyl tensor and the Gau\ss-Bonnet term.
(\ref{geomterms}) may be varied arbitrarily often with respect to $H$.
Thus the Ward identity for the superconformal current algebra is
obtained within the framework of general Green functions.  The terms
depending on $\be$ and $\gamma$ represent the dynamical, i.e. model
dependent anomalies of the superconformal currents, whereas the
r.h.s. corresponds to geometrical anomalies, in which only the
coefficients depend on the model.  With the help of these explicit
results it should be possible to discuss questions of the
superconformal current algebra in general.

\vspace{1cm}

\noindent
{\bf Acknowledgement}\\ We are grateful to Olivier Piguet for helpful
discussions on the BRS version of the $\Lambda$ transformations.

\newpage
\appendix

\section{Appendix  }

\renewcommand{\theequation}{\thesubsection.\arabic{equation}}

\setcounter{equation}{0}

\subsection{Fields and Transformation Laws}
\label{transformationlaws}

\begin{tabular}{|l|l|}
\hline Prepotentials & \tf{H = H^{\alpha\dot\alpha}
D_{\alpha\dot\alpha}} real \rule{0cm}{3ex}\\ & $\phi$
chiral\rule{0cm}{3ex}\\ & \tf{\phi=\e^J}, $J$ chiral \\ \hline
Vierbein determinant & \tf{E= \sdet \, E_A{}^M}
\raisebox{0ex}{\rule{0cm}{4ex}}\\ Curvature scalar & \tf{R = \bar D^2
\left(E^{-1} \phi^{-3}\right)}\raisebox{-3ex}{\rule{0cm}{7ex}}\\
\hline Matter field & $A$ chiral\raisebox{-1ex}{\rule{0cm}{4ex}} \\
\hline\hline $\Lambda$ transformations & \tf{\Lambda\phantom{_c} =
\Lambda^{\alpha\dot\alpha} D_{\alpha\dot\alpha} +
\Lambda^{\alpha}D_\alpha + \Lambda_{\dot\alpha} \bar D^{\dot\alpha}}
\rule{0cm}{3ex}\\ & \tf{\Lambda_c = \Lambda^{\alpha\dot\alpha}
D_{\alpha\dot\alpha} + \Lambda^{\alpha}D_\alpha}\rule{0cm}{3ex}\\ &
\tf{\Lambda^{\alpha\dot\alpha} = i \bar D^{\dot\alpha}
\Omega^\alpha}\rule{0cm}{4ex}\\ & \tf{\Lambda^\alpha = \frac{1}{4}
\bar D^2 \Omega^\alpha}\raisebox{-2ex}{\rule{0cm}{6ex}}\\ &
\tf{\Lambda_{\dot\alpha} = \e^{2iH} \bar
\Lambda_{\dot\alpha}}\rule{0cm}{3ex}\\ \cline{2-2} & \tf{\e^{2iH'} =
\e^\Lambda \e^{2iH} \e^{-\bar\Lambda}}\rule{0cm}{4ex}\\ &
\tf{{\phi'}^3 = \phi^3 \, \e^{\overleftarrow{\Lambda_c}}}
\rule{0cm}{3ex}\\ & \tf{J' = \e^{\Lambda} J + {\textstyle\frac{1}{3}}
\ln \left( 1 \cdot \e^{\overleftarrow{\Lambda_c}}\right)}
\rule{0cm}{3ex}\\ & \tf{(E\,')^{-1} = E^{-1} \,
\e^{\overleftarrow{\Lambda}}} \rule{0cm}{3ex}\\ & \tf{R' =
\e^{\Lambda} R \rule{0cm}{3ex}}\\ & \tf{A' = \e^{\Lambda} A
\rule{0cm}{3ex}}\\ \hline Super Weyl transformations & $\sigma$
chiral\rule{0cm}{3ex} \\ \cline{2-2} & $H'=H$\rule{0cm}{3ex}\\ &
\tf{\phi' = \e^\sigma \phi}\rule{0cm}{3ex}\\ &
\tf{J'=J+\sigma}\rule{0cm}{3ex}\\ & \tf{(E\,')^{-1} = E^{-1} \,\,
\e^\sigma \e^{2iH} \e^{\bar\sigma}}\rule{0cm}{3ex}\\ & \tf{R' =
\e^{-2\sigma} \left( \Db^2 + R\right)\e^{\bar\sigma}}\rule{0cm}{3ex}\\
& \tf{A' = \e^{-\sigma} A} \rule{0cm}{3ex}\\\hline
\end{tabular}

\subsection{Parameters of Superconformal Transformations}

\label{superconformalparameters}
\setcounter{equation}{0}
\begin{tabular}{|l|l|l|}
\hline Transformation & Parameter $\Omega^\alpha$ & Action on chiral
field $A$ \raisebox{-2ex}{\rule{0cm}{5ex}}\\ \hline General & $\bar
D^\da \Omega^\al = D^\al \bar \Omega^\da$, $\sigma=-\frac{1}{12}\bar
D^2 D^\al \Omega_\al$ & $\frac{1}{4} \bar D^2 (\Omega^\al D_\al A) -
\sigma A \,\,= \delta_\Omega A$ \raisebox{-3ex}{\rule{0cm}{7ex}}\\
\hline Translation & \tf{\Omega^\al = -\frac{i}{2}
{\sigma_a}^{\alpha\dot\alpha} t^a \bar\theta_{\dot\alpha}},
\,\,\tf{\sigma=0} & \tf{t^a \partial_a A \,\, = t^a \,\delta^P_a A}
\raisebox{-3ex}{\rule{0cm}{7ex}}\\ \hline Supersymmetry &
\tf{\Omega^\al = -\bar\theta^2 \, q^\alpha}, \,\,\tf{\sigma=0} &
\tf{q^\alpha (\partial_\alpha +i {\sigma^a}_{\alpha\dot\alpha}
\bar\theta^{\dot\alpha} \partial_a) A \,\, = q^\al \,\delta^Q_\al A}
\raisebox{-3ex}{\rule{0cm}{7ex}} \\ \cline{2-3} & \tf{\Omega^\al = 2
\theta^\alpha \bar \theta_\da \bar q^\da}, \,\,\tf{\sigma=0} &
\tf{\bar q_{\dot\alpha} ( -\bar\partial^{\dot\alpha} -i \theta^\alpha
{{\sigma^a}_\alpha}^{\dot\alpha} \partial_a) A \,\, = \bar q_\da
\,\delta^{\bar Q \da} A}\raisebox{-3ex}{\rule{0cm}{7ex}}\\ \hline $R$
Transformation & \tf{\Omega^\al = -i \theta^\alpha \bar\theta^2 \, r},
\,\,$\sigma=\frac{2}{3} i r$ & \tf{i r \, \left(-\frac{2}{3} +
\theta^\alpha \partial_\alpha + \bar\theta_{\dot\alpha}
\bar\partial^{\dot\alpha} \right)\, A \,\, = r \,\delta^R \! A}
\raisebox{-3ex}{\rule{0cm}{7ex}}\\ \hline Dilatation & \tf{\Omega^\al
= d \left(\half \theta^\alpha \bar\theta^2 -\frac{i}{2} x^a
{\sigma_a}^{\alpha\dot\alpha} \bar\theta_{\dot\alpha}\right)}, &
\tf{d\, \left(1 + x^a\partial_a +\half \theta^\alpha \partial_\alpha
-\half \bar\theta_{\dot\alpha} \bar \partial^{\dot\alpha}\right) A}
\raisebox{-3ex}{\rule{0cm}{7ex}}\\ & $\sigma=-d$ & \tf{ = d \,\delta^D
\!A}\raisebox{-2ex}{\rule{0cm}{5ex}} \\ \hline Lorentz &
\tf{\Omega^\al = i {\sigma_a}^{\al\da} \omega^{ab}x_b \bar\theta_\da},
\,\,$\sigma=0$ & \tf{\omega^{ab}\,\Bigl( x_a\partial_b - x_b\partial_a
-\frac{i}{2} \theta^\be {(\sigma_{ab})_\be}^\al \partial_\al}
\raisebox{-3ex}{\rule{0cm}{7ex}}\\ Transformation & & \quad \quad
\tf{+ \frac{i}{2} \bar \theta_\db {(\bar\sigma_{ab})^\db}_\da \bar
\partial^\da \Bigr) A \,\, = \omega^{ab} \delta^L_{ab} A}
\raisebox{-3ex}{\rule{0cm}{7ex}}\\ \hline S Transformation &
\tf{\Omega^\al = 2 s^\be x_a {\sigma^a}_{\be\db} \theta^\al
\bar\theta^\db}, & \tf{s^\al \, \Bigl( -x_a {\sigma^a}_{\al\da}
\delta^{\bar Q \da} +2 \theta_\al \delta^R - i\theta^2 D_\al}
\raisebox{-3ex}{\rule{0cm}{7ex}}\\ & $\sigma=4is^\al \theta_\al$ &
\quad\quad \tf{ -2i\left(d-\frac{n}{2}\right)\theta_\al \Bigr) A \,\,
= s^\al \,\delta^S_\al A} \raisebox{-3ex}{\rule{0cm}{7ex}}\\
\cline{2-3} & \tf{\Omega^\al = -\bar s_\da x_a \sigma^{a\al\da} \bar
\theta^2}, & \tf{\bar s_\da \Bigl( -x_a \delta^{Q\al}
{{\sigma^a}_\al}^\da + 2\bar\theta^\da
\delta^R}\raisebox{-3ex}{\rule{0cm}{7ex}} \\ & $\sigma=0$ & \quad\quad
\tf{+2i \left(d+\frac{n}{2}\right) \bar\theta^\da \Bigr) A \,\, = \bar
s_\da \,\delta^{\bar S \da} A} \raisebox{-3ex}{\rule{0cm}{7ex}}\\
\hline K Transformation & $\Omega^\al = \left(\frac{i}{2}k^b x^2 - i
k^a x_a x^b\right) {\sigma_b}^{\al\da} \bar \theta_{\da}$ & \tf{k^a \,
\Bigl( 2x_a \delta^D + 2x^b \delta^L_{ab} - 2x_ax^b\partial_b}
\raisebox{-3ex}{\rule{0cm}{7ex}} \\ & $+ k^a x_a \theta^\al
\bar\theta^2$, & \quad\quad \tf{+x^2 \partial_a + 2\theta
\sigma_a\bar\theta \delta^R -\frac{2}{3} i\theta\sigma_a\bar\theta}
\raisebox{-2ex}{\rule{0cm}{5ex}}\\ &
$\sigma=-2k^ax_a+2i\theta\sigma^a\bar\theta k_a$& \quad\quad \tf{ +
\theta^2 \bar\theta^2 \partial_a \Bigr)\, A \, = k^a \delta^K_a A}
\raisebox{-2ex}{\rule{0cm}{5ex}}\\ \hline
\end{tabular}

\subsection{Combined Diffeomorphism and Weyl Transformations}
\label{combinedtransformations}
Below we list the combined infinitesimal $\Lambda$ and $\sigma$
transformations in terms of $\Omega^\alpha$. In $\delta_\Omega H$ we
include only terms involving $\Omega$ but not $\bar \Omega$,
\begin{align}
 \delta_\Omega A &= \bar D^2 \left( {\ts \frac{1}{4}} \Omega^\alpha
 D_\alpha A \, - { \ts \frac{1}{12}} D_\alpha \Omega^\alpha A \right)
 \\[1ex] \delta_\Omega H^{\alpha\da} &= \half \bar D^\da \Omega^\alpha
 \nonumber\\ &\quad + \quar \bar D^{\dot\beta} \Omega^\beta
 \{D_\beta,\bar D_{\dot\beta} \} H^{\alpha\da} - \quar
 H^{\beta\dot\beta} \{D_\beta, \bar D_\db \} \bar D^\da \Omega^\alpha
 + \quar \bar D^2 \Omega^\beta D_\beta H^{\alpha\da}
                         \label{delta_Omega_H} \nonumber \\
      &\quad - {\ts \frac{1}{8}} H^{\gamma\dot\gamma} \{D_\gamma,\bar
D_{\dot\gamma}\} \bar D^2 \Omega^\be D_\be H^{\al\da} + {\ts
\frac{1}{24}} H^{\gamma\dot\gamma}\{D_\gamma,\bar D_{\dot\gamma}\}
\left( H^{\be\db} \{D_\be,\bar D_\db\} \bar D^\da \Omega^\al \right)
\nonumber \\ &\quad - {\ts \frac{1}{12}}
H^{\gamma\dot\gamma}\{D_\gamma,\bar D_{\dot\gamma}\} \left( \bar
D^\db\Omega^\be \{D_\beta,\bar D_{\dot\beta} \} H^{\al\da} \right)
\nonumber \\ &\quad +{ \ts \frac{1}{24}} \bar D^\db \Omega^\beta
\{D_\beta,\bar D_{\dot\beta} \} \left( H^{\gamma\dot\gamma}
\{D_\gamma,\bar D_{\dot\gamma}\} H^{\alpha\da} \right) \\ &\quad +
O(H^3) \nonumber\\[1ex] \delta_\Omega J &= \quar \bar D^2
(\Omega^\alpha D_\alpha J)
              \label{delta_Omega_J} \, .
\end{align}

\newpage
\subsection{Local Superconformal Ward Operators}
\label{wardoperators}
\setcounter{equation}{0} {\bf Diffeomorphisms ($\Lambda$
Transformations)}
\begin{align}
w_\alpha\il (A)&= \quar D_\alpha A \fdq{}{A}, \\[1ex] w_\alpha\il
(J)&= \quar D_\alpha J \fdq{}{J} - {\textstyle \frac{1}{12}} D_\al
\fdq{}{J},\\[1ex] w_\alpha\il (H)&=
{}w_\alpha^{(0)}(H)\,+\,w_\alpha^{(1)}(H)\,+\,w_\alpha^{(2)}(H)\,+\,O(H^3),\\[1ex]
w_\alpha^{(0)}(H) &= \half \bar D^\da \fdq{}{H^{\alpha\da}}, \\[1ex]
w_\alpha^{(1)}(H) &= \quar \bar D^\da \left( \{ D_\alpha,\bar D_\da \}
H^{\beta\db} \fdq{}{H^{\beta\db}} \right) +\quar \{D_\beta,\bar D_\db
\} \bar D^\da \left( H^{\beta\db} \fdq{}{H^{\alpha\da}}\right)
\nonumber\\ &\quad +\quar \bar D^2 \left( D_\alpha H^{\beta\db}
\fdq{}{H^{\beta\db}} \right),\\[1ex] w_\al^{(2)}(H) &= {\textstyle
\frac{1}{8}} \bar D^2 \{D_\gamma,\bar D_\dg\} \left(H^{\gamma\dg}
D_\al H^{\beta\db} \fdq{}{H^{\beta\db}}\right) \nonumber\\ &\quad
+{\textstyle\frac{1}{24}} \bar D^\da \{D_\gamma,\bar D_\dg\} \left(
H^{\gamma\dg} \{D_\beta,\bar D_\db\} \left(
H^{\beta\db}\fdq{}{H^{\al\da}}\right)\right) \nonumber\\ &\quad
+{\textstyle\frac{1}{12}} \bar D^\da \left(\{D_\al,\bar D_\da\}
H^{\beta\db} \{D_\gamma,\bar D_\dg\} \left(H^{\gamma\dg}
\fdq{}{H^{\beta\db}}\right)\right) \nonumber\\ & \quad +
{\textstyle\frac{1}{24}} \bar D^\da \left( \{D_\al,\bar D_\da\} \left(
H^{\gamma\dg} \{D_\gamma,\bar D_\dg\} H^{\beta\db} \right)
\fdq{}{H^{\beta\db}}\right).
\end{align}

{\bf Super Weyl Transformations}
\begin{equation}
\begin{array}{rclcrcl}
w\is(A) &=& -A\fdq{}{A}, & \qquad & w_\al\is(A) &=& -\frac{1}{12}
D_\alpha \left( A \fdq{}{A} \right), \\[1ex] w\is(J) &=& \fdq{}{J}, &
& w_\al\is(J) &= &\frac{1}{12} D_\al\fdq{}{J}, \\[1ex] w\is(H)&=&0 \,
, & & w_\al\is(H)&=&0 \, .
\end{array}
\end{equation}

{\bf Combined Transformations}
\begin{align}
w_\alpha (A)&= \quar D_\alpha A \fdq{}{A} -{\textstyle\frac{1}{12}}
D_\alpha \left( A \fdq{}{A} \right), \\[1ex] w_\alpha (J)&= \quar
D_\alpha J \fdq{}{J},\\[1ex] w_\alpha (H)&= w_\al\il(H).
\end{align}

\subsection{Expansion of the Action Functional}
\label{actionexpansion}
\setcounter{equation}{0}
\begin{align}
E^{-1} &= 1+ E^{-1\,(1)}+ E^{-1\,(2)}+ \dots\\ E^{-1\,(1)} &= {\ts
\frac{2}{3}} \bar D_\da D_\al H^{\al\da} + {\ts\frac{1}{3}} D_\al \bar
D_\da H^{\al\da} + J + \bar J\\ E^{-1\,(2)} &= {\ts\frac{2}{3}}
(J+\bar J)^2 \bar D_\da D_\al H^{\al\da} +{\ts\frac{1}{3}} (J+\bar J)
D_\al \bar D_\da H^{\al\da}\nonumber\\ &\quad +{\ts\frac{1}{2}}
(J+\bar J)^2 + H^{\al\da} \{D_\al,\bar D_\da\} \bar J \nonumber\\
&\quad + {\ts\frac{1}{6}} \bar D_\db \left(H^{\alpha\da} \{D_\alpha,
\bar D_\da \} D_\beta H^{\beta\db} \right) - {\ts\frac{1}{6}} \bar
D_\db \left( D_\be H^{\al\da} \{D_\al,\bar D_\da \} H^{\be\db} \right)
\nonumber \\ &\quad + {\ts\frac{1}{6}} \bar D_\db D_\be H^{\al\da}
\bar D_\da D_\al H^{\be\db} + {\ts\frac{1}{18}} \bar D_\db D_\be
H^{\be\db} \bar D_\da D_\al H^{\al\da} \nonumber\\ & \quad
+{\ts\frac{1}{18}} \{D_\be,\bar D_\db\} H^{\be\db} \{D_\al,\bar
D_\da\} H^{\al\da} +{\ts\frac{1}{6}} H^{\al\da} \{D_\al,\bar D_\da\}
\{D_\be,\bar D_\db\} H^{\be\db} \nonumber\\ &\quad + {\ts\frac{1}{9}}
\bar D_\db D_\be H^{\be\db} \{D_\al,\bar D_\da\} H^{\al\da}
\end{align}

\begin{align}
\Gamma &= \Gamma^{(0)}\,+\,\Gamma^{(1)}\,+\,\Gamma^{(2)}\,+\,
\dots,\\[3ex] \Gamma^{(0)} &= \frac{1}{16}\int \!\! \d^8z \,\, A
\,\bar A + \frac{g}{48} \int \!\! \d^6z \,\, A^3 + \frac{g}{48} \int
\!\! \d^6 \bar z \,\,\bar A^3 + \frac{m}{8} \int \!\! \d^6z \,\, A^2 +
\frac{m}{8}\int \!\! \d^6 \bar z \,\, \bar A^2 ,\nonumber\\[3ex]
\Gamma^{(1)} &= \frac{1}{8} \int \!\! \d^8z \,\,H^{\alpha\da}
V_{\alpha\da} + \left[ \frac{g}{16} \int\!\! \d^6z \,\,JA^3 +
\frac{3m}{8} \int\!\! \d^6z \,\,JA^2 +c.c. \right]\nonumber\\[1ex] &
\quad + \frac{1}{16} \int \!\! \d^8z (J+\bar J) A\bar A + \frac{1}{8}
\int \!\! \d^8z \left( \xi \bar J A^2+ \xi J\bar A^2\right)
\nonumber\\[3ex] \Gamma^{(2)} &= \frac{1}{16} \int \!\! \d^8z \Bigl[
-2AH^{\al\da}\{D_\al,\bar D_\da\} \left(H^{\be\db}\{D_\be,\bar D_\db\}
\bar A\right) + E^{-1\,(1)} AH^{\al\da}\{D_\al,\bar D_\da\}\bar
A\nonumber\\ & \quad\qquad\qquad + E^{-1\,(2)} A\bar A
\Bigr]\nonumber\\ &\quad + \frac{1}{8} \int \!\! \d^8z \Bigl[-2\xi
H^{\al\da}\{D_\al,\bar D_\da\} \left(H^{\be\db}\{D_\be,\bar D_\db\}
\bar A^2\right) +\xi E^{-1\,(1)} H^{\al\da}\{D_\al,\bar D_\da\}\bar
A^2\nonumber\\ & \quad\qquad\qquad + E^{-1\,(2)} (\xi A^2+\xi\bar
A^2)\Bigr] \nonumber\\ &\quad + \frac{9g}{24} \int\!\! \d^6z
\,\,J^2A^3 + \frac{9g}{24} \int\!\! \d^6\bar z \,\,\bar J^2\bar A^3 +
\frac{9m}{16} \int\!\! \d^6z \,\,J^2A^2 + \frac{9m}{16} \int\!\!
\d^6\bar z \,\,\bar J^2\bar A^2\nonumber
\end{align}

\subsection{$\bf \beta^\xi$ to order $\bf \hbar$}
\label{betafunction}
\setcounter{equation}{0}

To first order in $H$, zeroth order in $J, \bar J$, the supersymmetric
curvature scalar is given by
\begin{gather}
R \sim \, - {\ts \frac{2}{3}} i \, \bar D^2 \pr_a H^a \, ,
\end{gather}
such that to first order in $H$, the $R A^2$ term in the action
(\ref{WZcq}) generates the vertex in figure 1 while the kinetic term
generates the three vertices in figure 2.

\begin{center}
\begin{figure}[h]
\unitlength1cm \epsfxsize\textwidth \epsffile{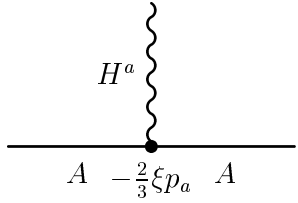}
\caption{Vertex with coupling $\xi$}
\end{figure}
\end{center}

\begin{center}
\begin{figure}[h]
\unitlength1cm \epsfxsize\textwidth \epsffile{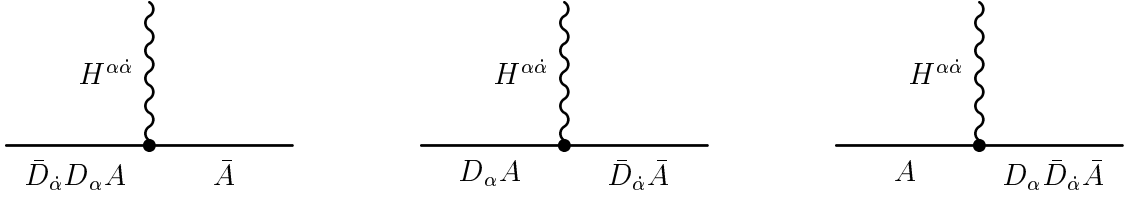}
\caption{Vertices for the kinetic term}
\end{figure}
\end{center}

Testing the CS equation (\ref{CSeq}) with respect to $H^a,A,A$ yields
to first order in $\hbar$
\begin{gather} \label{CS2}
\Gamma^{(1)}{}_{\! H^a AA} + \left( \beta^{\xi \, (1)} \pr_\xi +
 \beta^{g \, (1)} \pr_g - 2 \gamma^{(1)} \right) \Gamma
 ^{(0)}{}_{\! H^a AA} = 0
\end{gather}
at $s=1$.  The zeroth order vertex is given by
\begin{gather}
 \Gamma^{(0)}{}_{\! H^a{}_{\! 3} A_1A_2} = \, - {\ts \frac{2}{3}} \xi
p^H{}_{\! a} \delta^S(1,3,p) \delta^S(2,3,p+p^H) \, ,
\end{gather}
with $\delta^S$ the chiral $\delta$ function given by
\begin{gather}
\delta^S(1,2,p) = - {\ts \frac{1}{4}} \theta_{12}{}^{\!  2} \e^{- \bar
\theta_1 \gamma \theta_2 p} \, ,
\end{gather}
such that applying the normalisation condition (\ref{xinor}) to
(\ref{CS2}) we find
\begin{gather} \mu \pr_\mu \Big[
\frac{\pr}{\pr p^{H \,a}} \pr_{\theta_1}{}^{\!\! 2}
\pr_{\theta_2}{}^{\!\! 2} \Gamma^{(1) }_{H^a A A }(p_H ; p_1,p_2)
\Big]_{p^{H \,2}=p^2{}_{\! sym} = \mu^2 , \theta=\bar \theta= 0, s=1}
\, + \beta^{\xi \, (1)} - 2 \gamma^{(1)} = 0 \, , \label{CS1}
\end{gather}
with $p_1=p, p_2=p+ p^H$. For the subsequent calculation it is crucial
to note that the only terms in $\hat \Gamma_{HAA}$ contributing to the
first term in (\ref{CS1}) are those of the form $f(p^H)\theta_1{}^2
\theta_2{}^2$, with $f$ some function of $p^H$, and no further
$\theta, \bar \theta$ dependence.  The contributions to
$\Gamma^{(1)}{}_{\! HAA}$ are given by the graphs in figure 3 for the
$\xi$ dependent part and by figure 4 for the $\xi$ independent part.
All vertices of figure 2 contribute to the top vertex in both graphs
of figure 4.

\begin{center}
\begin{figure}[h]
\unitlength1cm \epsfxsize\textwidth \epsffile{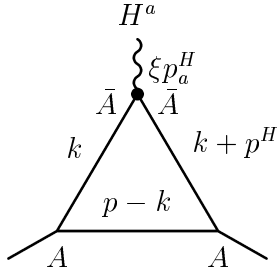}
\caption{$\xi$ dependent contribution to $\hat \Gamma_{H^a AA}$}
\end{figure}
\end{center}

\begin{center}
\begin{figure}[h]
\unitlength1cm \epsfxsize\textwidth \epsffile{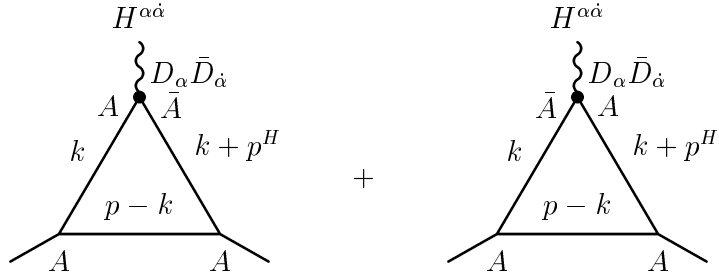}
\caption{$\xi$ independent contribution to $\hat \Gamma_{H^a AA}$}
\end{figure}
\end{center}

\noindent We note that according to (\ref{subdegree}), 
the subtraction degree is $\delta=1$.  For the propagators it is
useful to define
\begin{gather}
G_{1,2}(k,s) = \frac{i \delta^S(1,2,k)}{k^2 - M^2(s-1)^2 + i \eps} \,
\, .
\end{gather}
Then performing the subtraction of the integrand according to
(\ref{suboperator}) we find for the $\xi$ dependent part of the first
order vertex function, shown in figure 3,
\begin{gather} \label{gammaxi}
\Gamma^{\xi \, (1)}{}_{ H^a AA}(p^H,p) = {\ts \frac{2}{3}} \xi g^2 \,
\int \! \d^{\, 4} k \, \theta_{12}{}^{\! 2} \, p^H{}_{\! b} \Big[
\pr_{p_H}{}^{\! b} F_a(p^H, p, k, s=0) \Big]_{p^H=p=0} \, ,
\end{gather}
where
\begin{gather}
F_a(p^H,p,k,s) = 4 M (s-1) p^H{}_{\! a} G_{1,2}(p-k,s) D_1{}^{\! 2}
G_{1,3}(k,s) D_3{}^{\! 2} G_{3,2}(p^H+k,s) \, ,
\end{gather}
and the limit $s =1$ is taken at the end of the calculation. In
(\ref{gammaxi}) there is no contribution with a factor of $ p^H
\theta_1{}^{2} \theta_2{}^{2}$ without further $\bar \theta$
dependence such that (\ref{gammaxi}) does not contribute to the first
term of (\ref{CS1}).

When calculating the $\xi$ independent contributions to the first
order vertex function as shown in figure 4, we see that there are
potential contributions of the form $f(p^H) \theta_1{}^{\!2}
\theta_2{}^{\!2}$.  However these cancel when symmetrising the $H A
\bar A$ vertex, which corresponds to adding the two graphs in figure
4.  We have
\begin{align} 
\Gamma^{{\rm z} \, (1)}{}_{\! a \, HAA}(p^H,p) & = {\ts \frac{1}{3}}
 g^2 \, \int \! \d^{\, 4} k \, \, \Big( p^H{}_{\! a} \Big[
 \pr_{p_H}{}^{\! a} I(p^H, p, k, s=0) \Big]_{p^H=p=0} \nonumber\\ &
 \hspace{4cm} \label{gammaz} + p_{ a} \Big[ \pr_{p}{}^{\! a} I(p^H, p,
 k, s=0) \Big]_{p^H=p=0} \Big) \, ,
\end{align}
with
\begin{align}
I(p^H,p,k,s) =& M^2 (s-1)^2 G_{1,2} (p-k,s) \Big( \bar D_\da D_\al
G_{1,3}(k,s) D^2 G_{3,2}(k+p^H,s) \nonumber\\ & + D_\al G_{1,3}(k,s)
\bar D_\da D^2 G_{3,2}(k+p^H,s) - G_{1,3}(k,s) D_\al \bar D_\da D^2
G_{3,2}(k+p^H,s) \nonumber\\ & + D^2 G_{1,3}(k,s) \bar D_\da D_\al
G_{3,2}(k+p^H,s) + \bar D_\da D^2 G_{1,3}(k,s) D_\al G_{3,2}(k+p^H,s)
\nonumber\\ & \hspace{4cm} - D_\al \bar D_\da D^2 G_{1,3}(k,s)
G_{3,2}(k+p^H,s) \Big) \, , \label{otto}
\end{align}
where all derivatives act on the $(3)$ coordinates and $s=1$ is chosen
after performing all subtractions.  The first three terms in
(\ref{otto}) correspond to the first graph in figure 4 and the
last three to the second.  Using $\{ D_\al, \bar D_\da \}
\delta^S(1,2,p) = 2 \sigma^a_{\al \da} p_a \delta^S(1,2,p) $ it may be
checked that $\frac{\pr}{\pr p^{H \,a}} \pr_{\theta_1}{}^{\!\! 2}
\pr_{\theta_2}{}^{\!\! 2} \Gamma^{{\rm z} \, (1) }_{H^a A A } $
consists of a constant, momentum independent term.  Thus the first
term in (\ref{CS1}) vanishes and we have
\begin{gather} 
\beta^{\xi \, (1)} - 2 \gamma^{(1)} \xi = 0 \, .
\end{gather}

\end{document}